\documentclass[pdflatex,sn-mathphys-num]{sn-jnl}

\usepackage{graphicx}%
\usepackage{multirow}%
\usepackage{amsmath,amssymb,amsfonts}%
\usepackage{amsthm}%
\usepackage{mathrsfs}%
\usepackage[title]{appendix}%
\usepackage{xcolor}%
\usepackage{textcomp}%
\usepackage{manyfoot}%
\usepackage{booktabs}%
\usepackage{algorithm}%
\usepackage{algorithmicx}%
\usepackage{algpseudocode}%
\usepackage{listings}%

\usepackage{graphicx, bm}

\usepackage{graphicx}
\usepackage{tikz}
\usetikzlibrary{arrows, shapes, positioning, calc}
\usepackage{quantikz}

\theoremstyle{thmstyleone}%
\theoremstyle{thmstyletwo}%

\theoremstyle{thmstylethree}%

\begin{document}


\title{ Generative Adversarial Variational Quantum Kolmogorov-Arnold Network }

\author*[1]{\fnm{ Hikaru } \sur{ Wakaura }}\email{ hikaruwakaura@gmail.com }

 \affil[1]{ QuantScape Inc. QuantScape Inc., 4-11-18, Manshon-Shimizudai, Meguro, Tokyo, 153-0064, Japan }

  
\abstract{ Kolmogorov–Arnold Networks (KANs) are a recently proposed class of multi-layer neuromorphic networks that achieve higher approximation accuracy than conventional neural networks with fewer trainable parameters. Due to these advantages, KANs have attracted increasing attention and have been applied to various learning tasks. Although KANs can, in principle, be used as generative models either independently or within a Generative Adversarial Network (GAN) framework, their training cost scales poorly with the number of parameters, resulting in slower learning compared to standard neural networks. Consequently, KAN-based models have seen limited exploration in generative learning.

In this work, we propose a novel generative framework, Generative Adversarial Variational Quantum Kolmogorov–Arnold Network (GAVQKAN), which employs a Variational Quantum KAN as the generator. By leveraging the expressive power of parameterized quantum circuits and their probabilistic output distributions, the proposed method enables efficient generative learning with significantly fewer parameters. We evaluate GAVQKAN on the MNIST and CIFAR-10 datasets and compare its performance with classical neural networks and Quantum Generative Adversarial Networks. Experimental results demonstrate that GAVQKAN achieves higher generation accuracy while requiring fewer training samples, highlighting its potential as a scalable quantum-enhanced generative model. }

\keywords{ Quantum machine learning, Kolmogorov-Arnold Network, Variational Quantum Algorithms}

  
 
\maketitle

\section{Introduction}\label{1}

Quantum computing has long been anticipated to enable the efficient solution of problems that are intractable for classical computers, a vision first articulated by Feynman in 1982 \cite{feynman_simulating_1982}. 
Since then, several quantum algorithms have demonstrated theoretical or empirical advantages over classical approaches, including Grover's search algorithm \cite{2003quant.ph…1079L}, the quantum Fourier transform \cite{PRXQuantum.4.040318} and its applications \cite{2024arXiv241004435I,2023arXiv230504908M}, and Shor's factoring algorithm \cite{2023arXiv230609122S}. 
These developments highlight the potential of quantum computation to surpass classical limits in specific problem domains.

Despite this promise, the realization of large-scale quantum computation remains constrained by hardware limitations. Current quantum processors are highly susceptible to noise, and fully fault-tolerant architectures are not yet available. Consequently, most existing devices operate in the Noisy Intermediate-Scale Quantum (NISQ) regime, where circuit depth and gate fidelity are severely limited. These constraints necessitate algorithmic frameworks that can tolerate noise while remaining expressive enough to address practically relevant problems. 
      
Variational quantum algorithms (VQAs) have emerged as a leading paradigm for computation on NISQ devices. By combining parameterized quantum circuits with classical optimization loops, VQAs reduce circuit depth requirements and shift complexity to classical post-processing. This hybrid structure has enabled a wide range of applications, particularly in quantum machine learning, where variational models have been explored for classification, regression, and generative tasks under realistic hardware constraints.  
       
Recently, Kolmogorov–Arnold Networks (KANs)\cite{2024arXiv241106727C,2024arXiv241118165H,2024arXiv241203710K,2024arXiv241008452B,2024arXiv241114902K,2024arXiv241007446J,2024arXiv240800273T} have attracted increasing attention as an alternative to conventional neural networks. 
Unlike standard architectures that rely on linear combinations of fixed activation functions, KANs employ learnable univariate functions, offering enhanced function approximation capabilities. Several quantum extensions of KANs have been proposed, including fault-tolerant implementations based on block encoding and variational formulations compatible with NISQ hardware. These quantum KAN models have been reported to outperform conventional quantum neural networks in terms of accuracy and expressivity.  
     
However, existing quantum KAN approaches face a fundamental scalability challenge. The number of trainable parameters scales with the digital length of the output data, similarly to classical neural networks, and the computational cost grows quadratically with respect to the parameter count. This limitation significantly restricts the applicability of quantum KAN families to complex or high-dimensional generative tasks, where large output representations are unavoidable.  
The Variational Quantum KANs have the potential to go across the hardles because they can generate the $ 2 ^ { N _ q } $-dimensional data for the number of qubits $ N _ q $ how few the number of parameters is by the quantum state on quantum computers.

In this work, we propose Generative Adversarial Variational Quantum Kolmogorov–Arnold Networks (GAVQKAN), a framework designed to overcome this scalability bottleneck. GAVQKAN integrates Variational Quantum KANs with a generative adversarial learning scheme and a batch-based training strategy, enabling the generation of high-dimensional data without increasing the number of variational parameters. This design allows quantum KAN-based models to scale to more complex data distributions while remaining compatible with NISQ-era hardware.
         
We evaluate GAVQKAN on the MNIST, CIFAR-10, and Fashion-MNIST datasets and compare its performance with Quantum Generative Adversarial Networks, classical neural networks, and Quantum Reservoir GANs. Our experimental results demonstrate that GAVQKAN achieves higher generation fidelity while requiring fewer training data than both quantum and classical baseline models.
   
The remainder of this paper is organized as follows. Section~\ref{1} introduces the background, Section~\ref{2} presents the proposed GAVQKAN framework, Section~\ref{3} reports experimental results, Section~\ref{5} discusses the implications of our findings, and Section~\ref{7} concludes the paper.
  
\section{ Related Works}
 
Variational Quantum Algorithms on NISQ Devices :
The limitations of current quantum hardware have motivated extensive research on variational quantum algorithms (VQAs), which are designed to operate effectively in the Noisy Intermediate-Scale Quantum (NISQ) regime. Early work demonstrated that hybrid quantum–classical optimization can mitigate circuit depth constraints while maintaining expressive power \cite{McClean_2016}. Subsequent developments include Adaptive VQE \cite{Grimsley2019}, Multiscale Contracted VQE (MCVQE) \cite{2019arXiv190608728P}, and various extensions tailored to specific problem domains \cite{2021arXiv210501141W,2021arXiv210902009W}. These approaches establish VQAs as a practical paradigm for near-term quantum computation. 

Variational Quantum Algorithms for Quantum Machine Learning :
  Building upon the VQA framework, numerous studies have explored quantum machine learning (QML) models that leverage parameterized quantum circuits for learning tasks \cite{2014PhRvL.113m0503R,2019QS&T....4a4001K,2019Natur.567..209H}. In particular, generative models such as Quantum Generative Adversarial Networks (QGANs) have been proposed to enable data generation and distribution learning on quantum hardware \cite{2022PhRvA.106b2601A,2022arXiv220211200K,2021PhRvP..16d4057B}. While these models demonstrate the feasibility of quantum-assisted generative learning, their scalability remains constrained by circuit depth, parameter count, and training stability.
 
Quantum Neural Networks :
Quantum neural networks (QNNs) represent a broad class of models that mimic classical neural network structures using quantum circuits \cite{PhysRevA.98.032309}. Various architectures have been proposed, including layered ansätze and data re-uploading schemes \cite{2020PhRvL.125j0401W}. Although QNNs have shown promise for classification and generative tasks, their expressivity and trainability often degrade as model complexity increases, particularly under NISQ constraints.
   
Quantum Kolmogorov–Arnold Networks :   
Recently, Kolmogorov–Arnold Networks (KANs) have been introduced as an alternative to conventional neural networks, offering improved function approximation through learnable univariate functions. Several quantum extensions of KANs have been proposed. Quantum KAN employs block encoding to realize KAN architectures on fault-tolerant quantum computers \cite{2024arXiv241004435I}. Variational Quantum KAN (VQKAN) adapts KANs to the VQA framework, enabling execution on NISQ devices \cite{Wakaura_VQKAN_2024}. In addition, KAN-based approaches have been applied to quantum architecture search \cite{2024EPJQT..11...76K}. While these methods demonstrate improved accuracy compared to conventional QNNs, they inherit a critical limitation: the number of trainable parameters scales with the digital length of the output, and the computational cost grows quadratically with respect to the parameter count \cite{Wakaura_VQKAN_2024}. This scalability issue motivates the development of alternative training and generation strategies for quantum KAN-based models.

\section{Method}\label{2}  
In this section, we describe the details of GAVQKAN.  
 
GAVQKAN is the Quantum GAN that uses KAN on ansatz and generates the data obeying the manner of Born Machine.  
GAN is one of the generative models that combines two networks or models in machine learning.   
The discriminator distinguishes genuine data from fake data generated by the generator network, and the generator network makes the fake data pass the discriminator's judgment as genuine data.      
Both are optimized to accomplish the assigned task, and they are expressed as an equation, 
\begin{equation}   
\label{eq_gan}   
\min_G \max_D V_{\text{\tiny GAN}}(D, G) = \mathbb{E}_{\bm{x} \sim p_{\text{data}}(\bm{x})}[\log D(\bm{x})] + \mathbb{E}_{\bm{z} \sim p_{\bm{z}}(\bm{z})}[\log (1- D(G(\bm{z})))] 
\end{equation}
Then, $ D(\bm{x}) $ is the probability that the data $ \bm{x} $ is distinguished correctly for the boolean label of the data $ p_{\text{data}}(\bm{x}) $ on discriminator $ D $, and $ G(\bm{z}) $ is the data made by generator $ G $ from latent vector $ \bm{z} $, respectively.
The boolean label of the data $ \sim p_{\text{data}}(\bm{x}) $ becomes 0 if the data is fake and 1 if genuine, and $ p_{\bm{z}}(\bm{z}) $ is 1 for all cases.
In simple form, it is expressed as,
\begin{eqnarray}  
V_{\text{\tiny GAN}}(D, G) &=& L _D + L _G \\
L _D &=& p \log D(\bm{x}) + (1- p) \log D(G(\bm{z})) \\
L _G &=&- p \log (1- D(G(\bm{z}))) \\\nonumber
\end{eqnarray}  
which $ p $ is the boolean label.
They are loss functions to be minimized, hence, the generator loss function is referred as 
\begin{equation} 
l_G = -  \log (D(G(\bm{z}))), 
\end{equation}

to be maximized.

\begin{figure}          
\centering    
  
 \includegraphics[scale= 0.2 ]{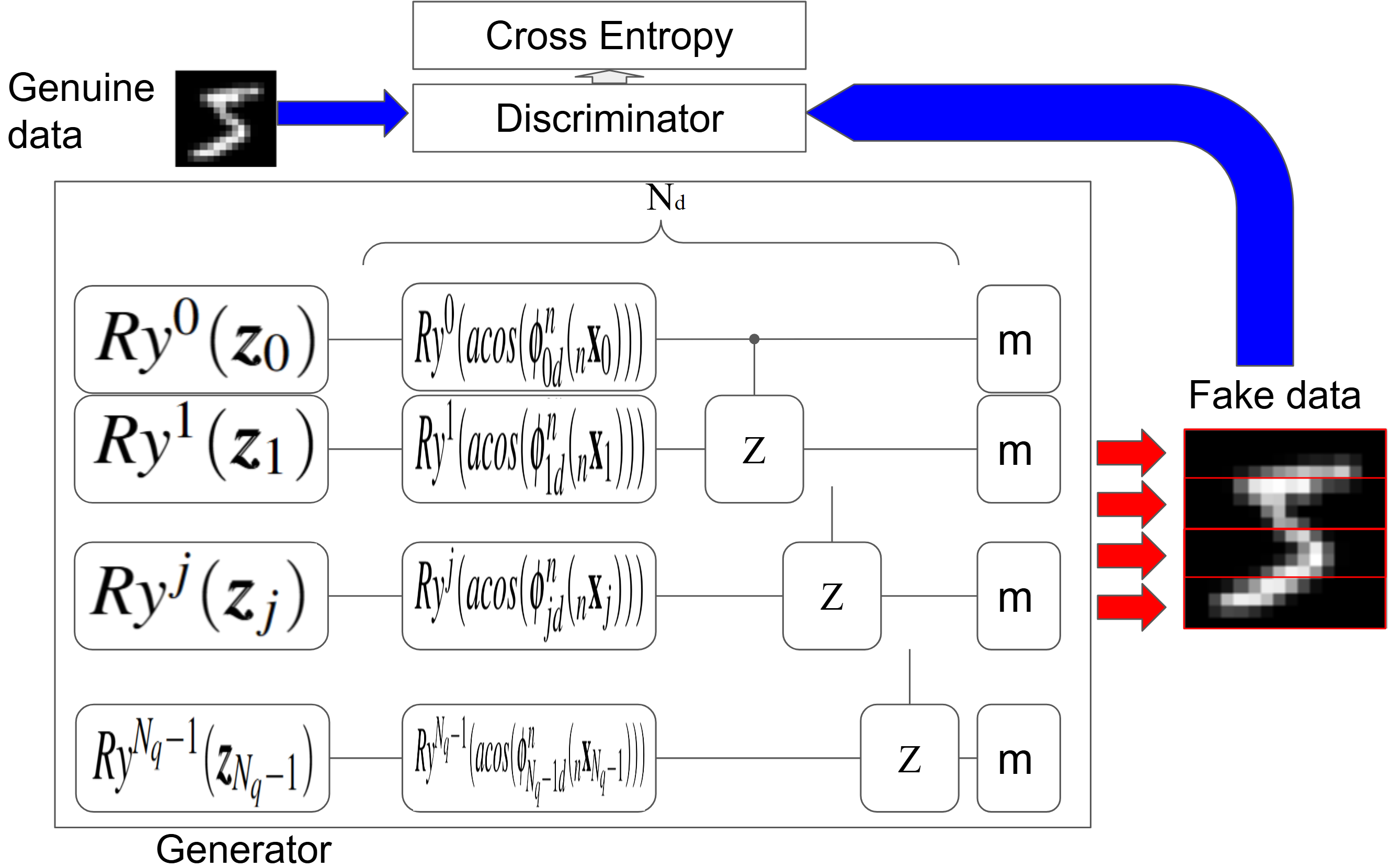} 
  
\caption{ The simplified illustration of our method.   
 Each qubit is initialized by a latent vector $ z $.  
Qubits are calculated by Ry gates, and blended by control - Z gate, respectively by layer.         
 Quantum circuits emerge from probability distributions as patches, and they are unified into one data.   
Loss functions are two cross entropy calculated from genuine and generated data. } \label{ g a v q }      
     
\end{figure}

We generate the $ G(\bm{z}) $ by the Variational Quantum KAN using the latent vector $ z$ as noise.
VQKAN is the variational quantum algorithm version of KAN, a multi-layer network based on the connection of synapses in neurons.    
 First, initial state $ \mid \Psi_{ini} ( \bm{z} ) \rangle $ is $ \prod _{j = 0} ^{N _q -1 } Ry^j ( \bm{z} ) \mid 0 \rangle ^{\otimes N _q} $ for each input $ m $.  
$ Ry^{j} (\theta) $ is $ \theta $ degrees angle rotation gate for y-axis on qubit $ j $.  
$ _n {\bf x}  $ is the input vector at layer n which $ _1 {\bf x} = \bm{z} $.  
For VQKAN, $ \phi_{j d }^{n} (_n {\bf x} ) $ is the gate of the angle.      
              
\begin{equation}      
\phi_{j d }^{n } (_n {\bf x} ) = \sum_{i \in \{0, dim (_n {\bf x} ) \}}^{ 0 } 2 acos (E_f (_n x_i ) +\sum_{s = 0 }^{N_g -1 }\sum_{l = 0 }^{N_s -1 } c_s^{n j d } B_l (_n x_i ))   \label { k a n }.  
 \end{equation}            
, which $ n $ is the index of the layer, $ N_g $ is the number of grids for each gate, $ N_s $ is the number of splines, respectively.       
 Then, $ c_s^{n j d } $ and $ B_l (_n x_i ) $ are the parameters to be trained, initialized into 0 and B spline functions at layer n whose domains are $ [  0, 1 ] $, respectively, the same as classical KAN.
The number of parameters is $ N _ q N _ d N _ l N _ g $. 
$ _n {\bf x}  $ is the input vector at layer n, $ d $ is the index of the depth in the single layer, and $ j $ is the index of qubits, respectively.  
 $ E_f (_n x_i ) = _n x_i  / (exp (-_n x_i ) + 1) $ is the Fermi-Dirac expectation energy-like value of the distribution. The component of $ _n {\bf x} $ is the expectation value of the given observable for the calculated states of qubits.    
The layers are the combined of para metric gates called ansatz.  
In detail the single layer $ \Phi _n ^G = \prod _{ d = 0} ^{N _ d -1 } (  \prod _{j = 0} ^{N _q -1 } Ry^j ( \phi_{j d  }^{n } (_n {\bf x} _j ) ) \prod _{j = 0} ^{N _q - 2 } CZ_{ j, j + 1 } ) $ as shown in Fig.\ref { g a v q } with the abstract picture of the flow of GAVQKAN. 
The entire ansatz is as follows.                    
\begin{equation}      
\mid \Psi ( _1 {\bf x}  ) \rangle = \prod_{ n = 1 }^{ num. ~ of ~ layers ~ N_l} { \Phi}_n^{ G }M \mid \Psi_{ ini } ( _1 {\bf x}  ) \rangle     
\end{equation} 
    
Then, $ M $ indicates the measurement of all qubits and deriving distribution of probability as a $ 2 ^ { N _ q } $ - length vector at a maximum length  to make $ _ { n + 1 }  {\bf x} $.  
$ _n {\bf x} _j $ for $ n > 1 $ is the average of the $ N_q $ equally separated segments of output data.

  The general quantum machine learning algorithms optimize ansatz which make initial states determined by latent vector $ \bm{z} ^ j $ aimed states, hence, one iteration of parameters $ \bm { \theta } $ is expressed asuming the steepest gradient decend as follows, 
\begin{equation}
\bm { \theta } ^ { j + 1 }= \bm { \theta } ^ j - ( f _ { aim } - f ( \bm { \theta } ^ j ) ) / \delta \bm { \theta }. 
\end{equation}
The distance and angle of each data from initial states are all different from others, hence, $ \bm { \theta } $s approach the center of mass which zero point is initial state. 
The broader aimed states scatter, the further some aimed states be from the center of mass. 
In contrast, VQKAN's ansatz depends on the latent vectors, hence, one iteration of parameters function $ \bm { \phi } $ is expressed asuming the steepest gradient decend as follows, 
\begin{equation}
\bm { \phi } ^ { j + 1 } ( \bm { z } ^ { j + 1 } ) = \bm { \phi } ^ j ( \bm { z } ^ { j + 1 } ) - ( f _ { aim } - f ( \bm { \phi } ^ j ( \bm { z } ^ { j + 1 } ) ) ) / \delta \bm { c }. 
\end{equation}
VQKAN's ansatse make initial states near or exact aimed states corespond to their latent vectors, hence, the ansatse optimize projection of each initial state to corresponding aimed state. 
Hence, GAVQKAN is able to store more conbinations of latent vector and aimed state than conventional Quantum GAN.

The condition of convergence is the default of scipy for all methods.   
We assume $ N_l = 1, N_q =  8, N_g = 8 $ and $ N_s = 4 ( tr + 2 ) $ for the number of trials $ tr $, respectively. 
$ G(\bm{z}) $ is the combined probability distribution from 4 patches.
The length of probability distribution we use as the patch is the quarter of the length of the entire image as a vector.  
We use stochastic gradient descent as an optimizer on both the discriminator and the generator, and the learning rates are 0.1 and 0.001, respectively.
We train once on an iteration on one image. 
The number of  samples to process on the GAVQKAN of an iteration is fixed to 1. 

All calculations are performed using PyTorch and Google Colab.  
The included programs are as follows; Python : 3.12.12, NumPy : 2.0.2, torch : 2.9.0, swd : 1.0.0, and  PennyLane : 0.43.1. 
We use PennyLane as a quantum simulator.  
 
\section{Result of numerical simulations}\label{3}
 
 In this section, we describe the results of generating the pictures on the MNIST and CIFAR-10 dataset.
  First, we compared the results on GAVQKAN, Quantum GAN (QGAN), and Classical Neural Network (CNN) using loss functions of cross-entropy on 16 $ \times $ 16-sized handwritten single numbers for 1000 samples, respectively.    
QGAN \cite{2021PhRvP..16b4051H} is simulated by PennyLane, which number of qubit is 8 and the depth is 6, and ansatz is $ \prod _{ j = 0 } ^{ N_q- 1 } R y (z_j) \prod _{ k = 0 } ^{ N _ d - 1 } \prod _{ j = 0 } ^{ N_q- 1 } R y (\theta_ { j, k }) \prod _{ j = 0 } ^{ N_q- 2 } C Z _{ j, j + 1 } $ for noise and $ N _ q N _ d $ parameters for each qubit. 
 CNN is performed by a 6-layer network consisting of layers sized   [ 256, 1024, 2048, 6272, 6272, 256 ].   
Discriminator is CNN consists of 3 layers sized [ $ N_{ data } $, $  2 \sqrt { N_{ data } }  $, 1 ]  for the data length $ N_{ data } $ with the rectified linear unit function between marginal layers and Sigmoid at the end.

\begin{figure}    

\includegraphics[scale=0.3]{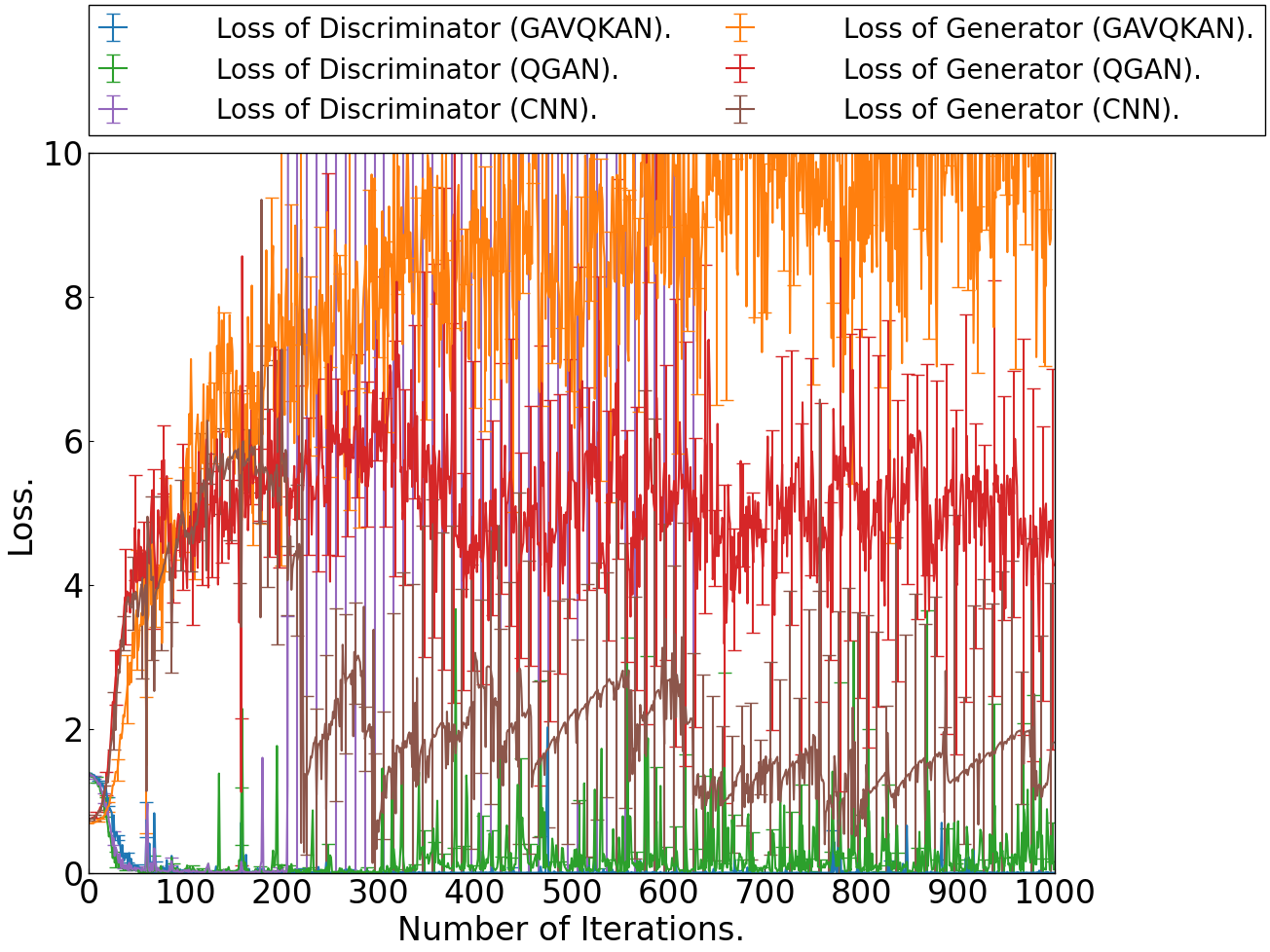}

\caption{ The number of iterations v.s. the average of the loss function for 10 attempts of the discriminator $ L_D$ and generator $ l_G$ of GAVQKAN, QGAN, and CNN using  cross-entropy as a loss function.
Error bars indicate the standard deviation from the average, and each bar is sampled every 10 points, moving 2 points from the previous data.
} \label{ comp 0 n }   
\end{figure}

\begin{figure} 
\includegraphics[scale= 0.2]{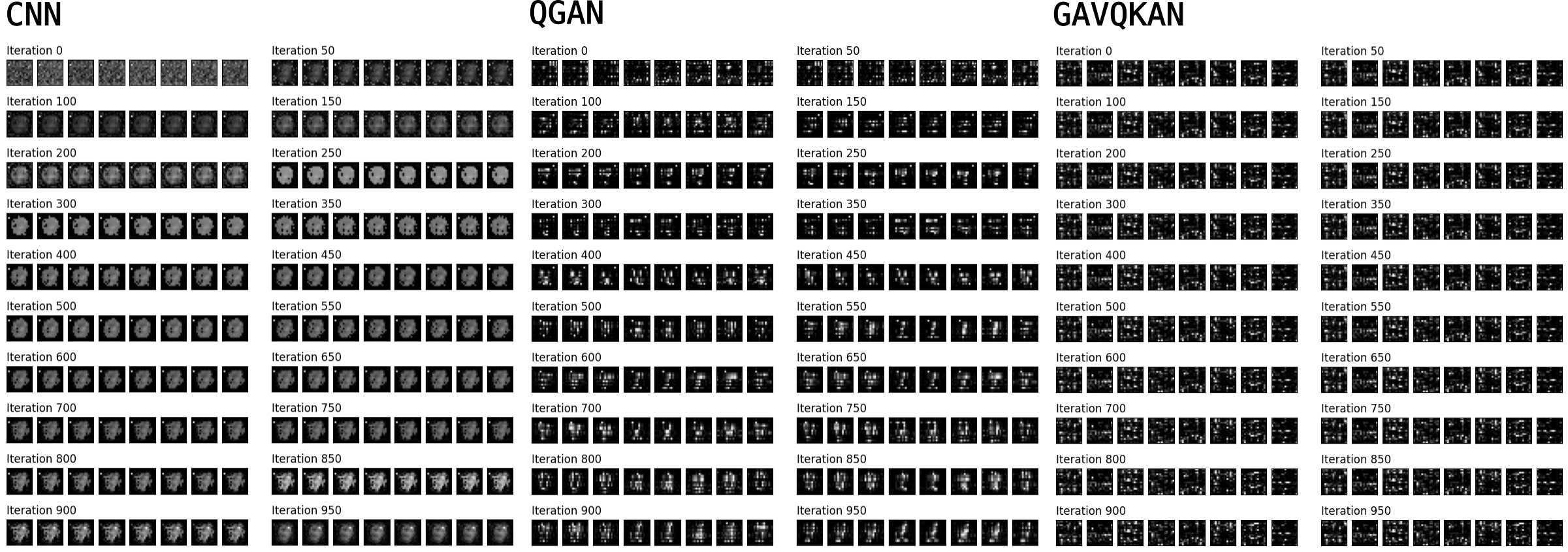}
\caption{ The generated picture of handwritten numbers by GAVQKAN, QGAN, and CNN using cross-entropy as a loss function for the number of iterations.
The picture by GAVQKAN is generated by fixed noise and input.  } \label{ gen 0 n }
\end{figure}

We show the average of the loss function for 10 attempts of the discriminator and generator of GAVQKAN, QGAN, and CNN for the number of epochs using cross-entropy in Fig.\ref{ comp 0 n }, and the generated picture of 0 in Fig.\ref{ gen 0 n }, respectively.    
The loss functions of generator GAVQKAN became over 6 the firstest, and QGAN is the second firstest.  
In contrast, the loss function of the discriminator of GAVQKAN became 0 the slowest in all, and QGAN is the second slowest.
The generated pictures of GAVQKAN are close to genuine data, even though the number of iterations is under 100.    
Although, generated pictures of QGAN became closer to genuine data than those of  GAVQKAN when the number of iterations is over 400.
Sliced Wasserstein Distance (SWD) also exhibits the accuracy and their tendency.
\begin{figure}   

\includegraphics[scale=0.3]{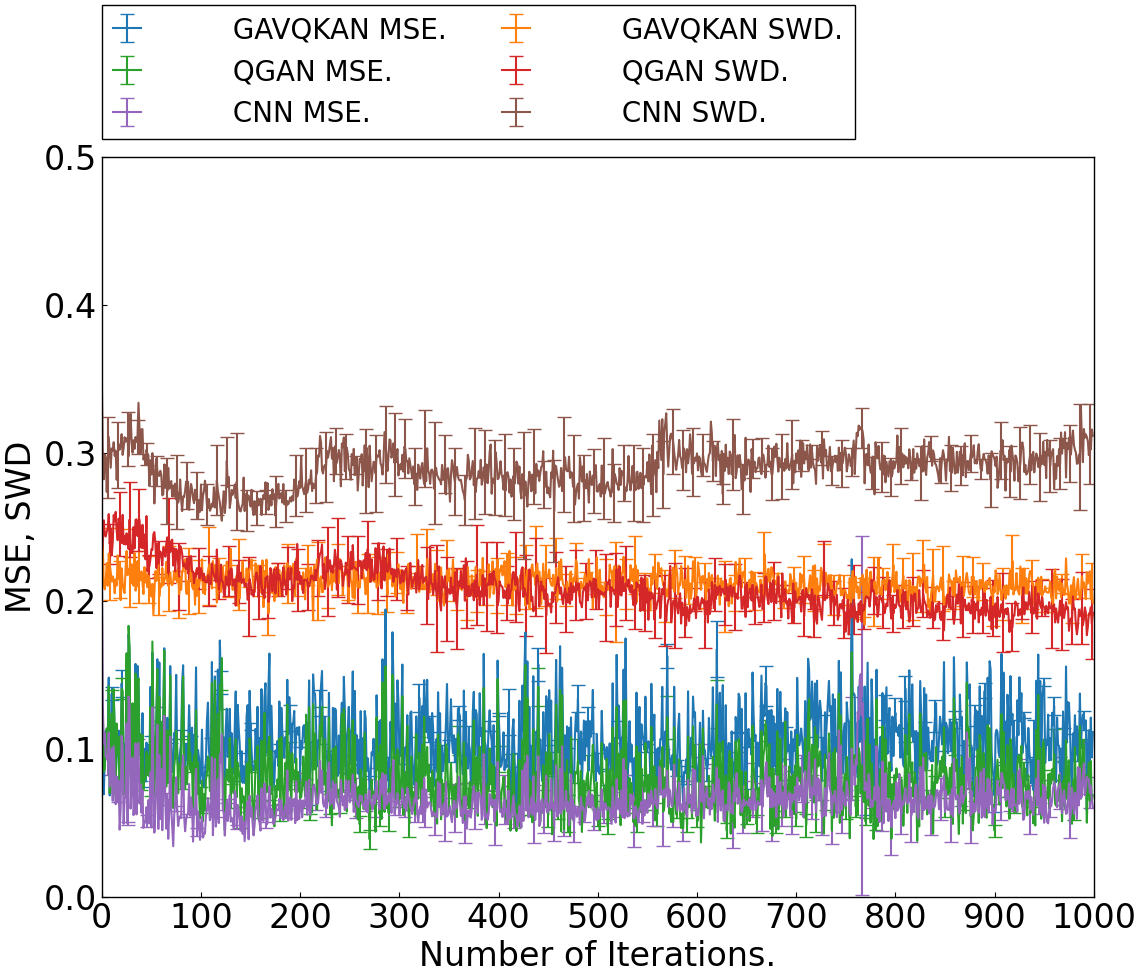}

\caption{ The number of iterations v.s. the average of the MSE and SWD for 5 attempts of GAVQKAN, QGAN, and CNN using cross-entropy as a loss function on the training and generating the MNIST dataset. } \label{ w d a }
\end{figure}

\begin{figure}  
   
\includegraphics[scale=0.3]{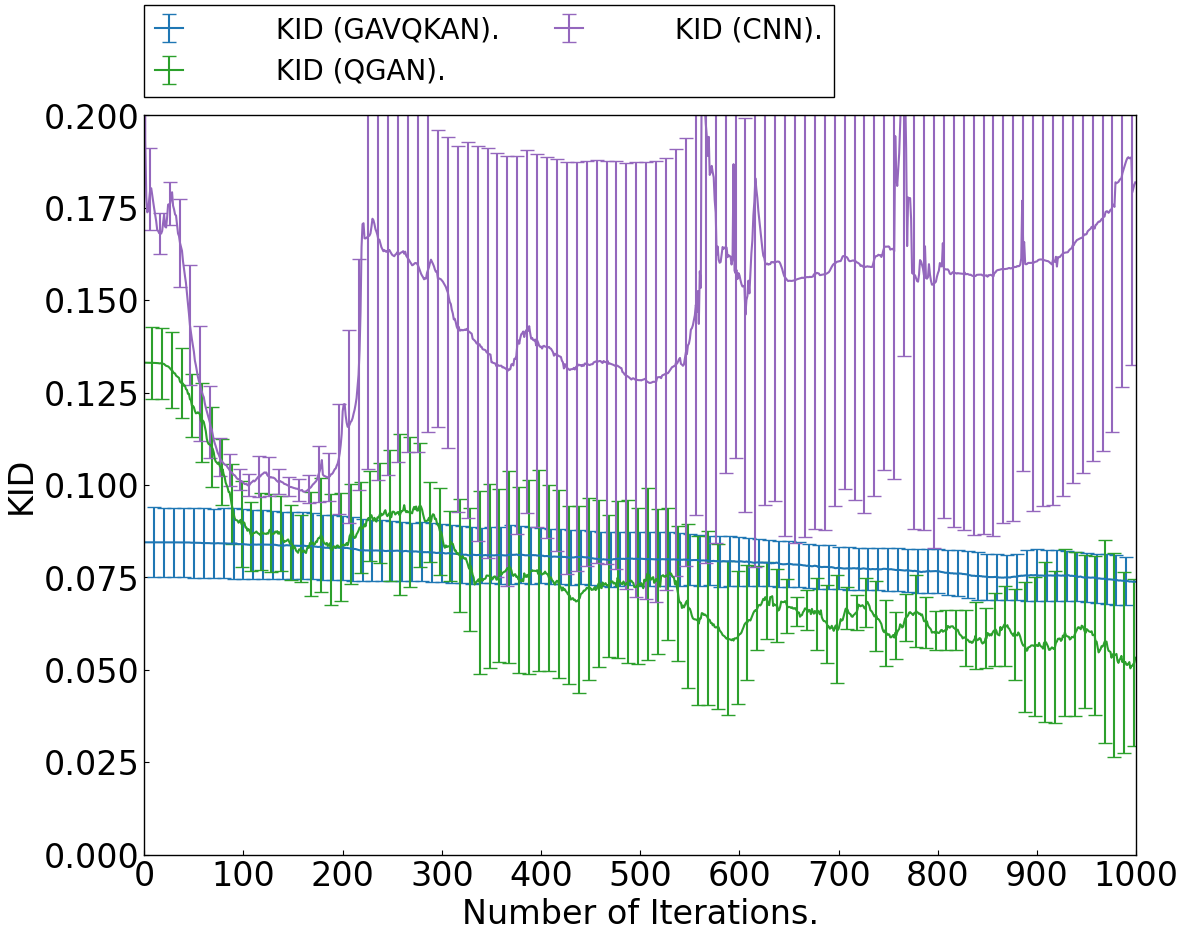}  

\caption{ The number of iterations v.s. the average of the KID for 5 attempts of GAVQKAN, QGAN, and CNN using cross-entropy as a loss function on the training and generating the MNIST dataset. } \label{ d a }

\end{figure} 

We also show the Mean Squared Error ( MSE ) and SWD of three methods of the first 8 pictures for fixed noise in Fig.\ref{ w d a }.     

SWDs are calculated by Python Optimal Transport \cite{JMLR:v22:20-451} on the first 8 pictures in the index of the original dataset.        
SWD of GAVQKAN in range the number of iterations below 300 is the smallest in all, and the second smallest over 300.    
In addition, MSE of GAVQKAN is on each learned data is the largest in all on all range.
We also show the Kernel Inception Distances ( KID ) of three methods of the first 8 pictures for fixed noise in Fig.\ref{ d a }.  
KID of GAVQKAN is the smallest in the range; the number of iterations below 300 is the smallest, the same as SWD.    
It lowers gradually in all ranges different for others because the KID of each attempt of other methods saturates in the range over 900.    
GAVQKAN can generate the data accurately by the smallest numbers of data in  GAVQKAN, QGAN, and CNN.            
However, the accuracy itself is not high compared to QGAN.        

The loss function of the discriminator of GAVQKAN has the smallest standard deviation in all, hence, learning ratio is supposed not to be optimum values.    
Besides, the time for calculations is 5.5952 times larger than QGAN and 10.1666 times larger than CNN, respectively.   
The number of parameters and ansatz must be surveyed further to improve both calculation time and accuracy.

\begin{figure}

\includegraphics[scale=0.3]{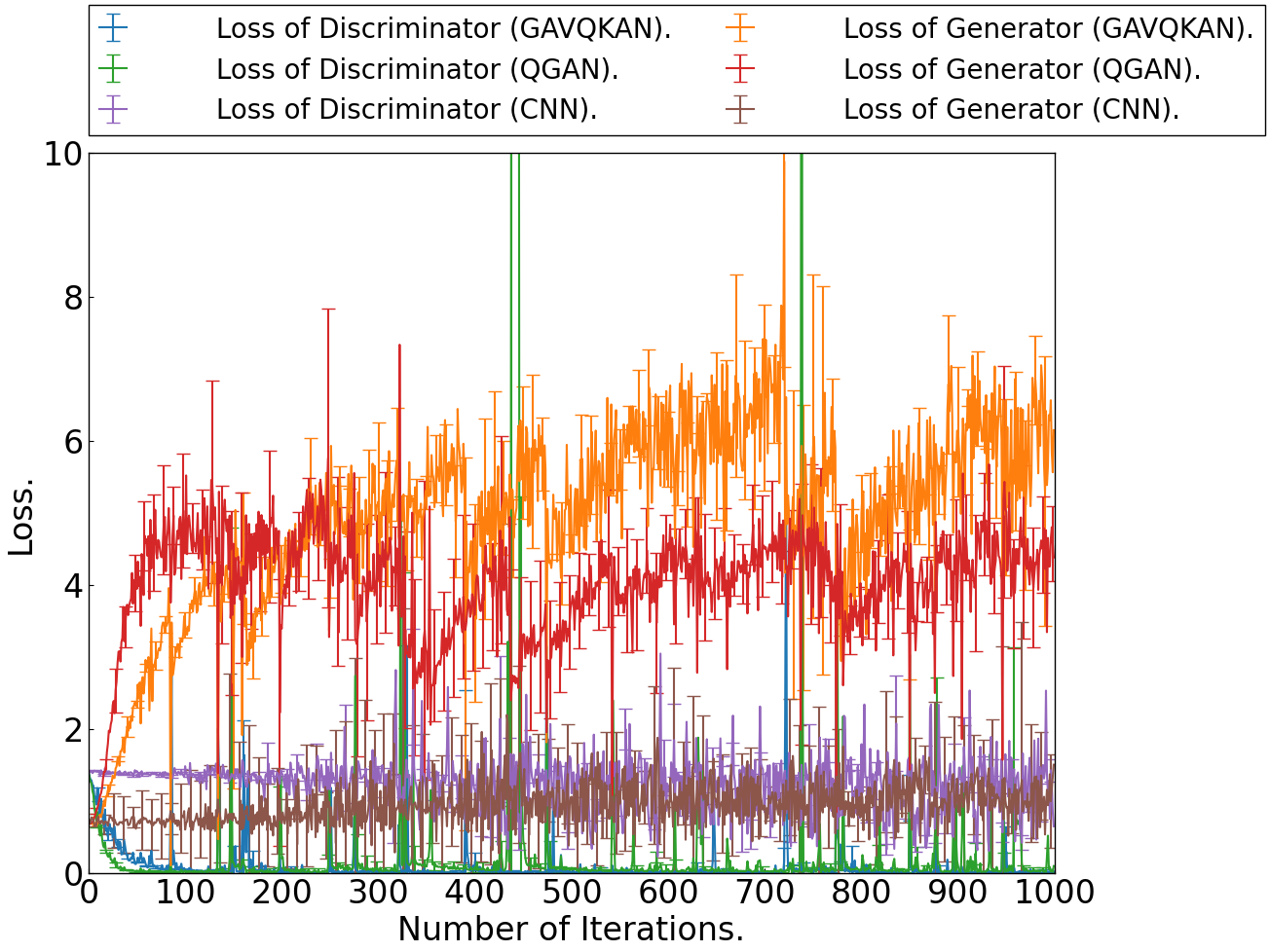}
 
\caption{
The number of iterations v.s. the average of the loss function for 5 attempts of the discriminator and generator of GAVQKAN, QGAN, and CNN using cross-entropy as a loss function on the training and generating the CIFAR-10 dataset.} \label{ pic s }
\end{figure}

\begin{figure}   
\includegraphics[scale= 0.2]{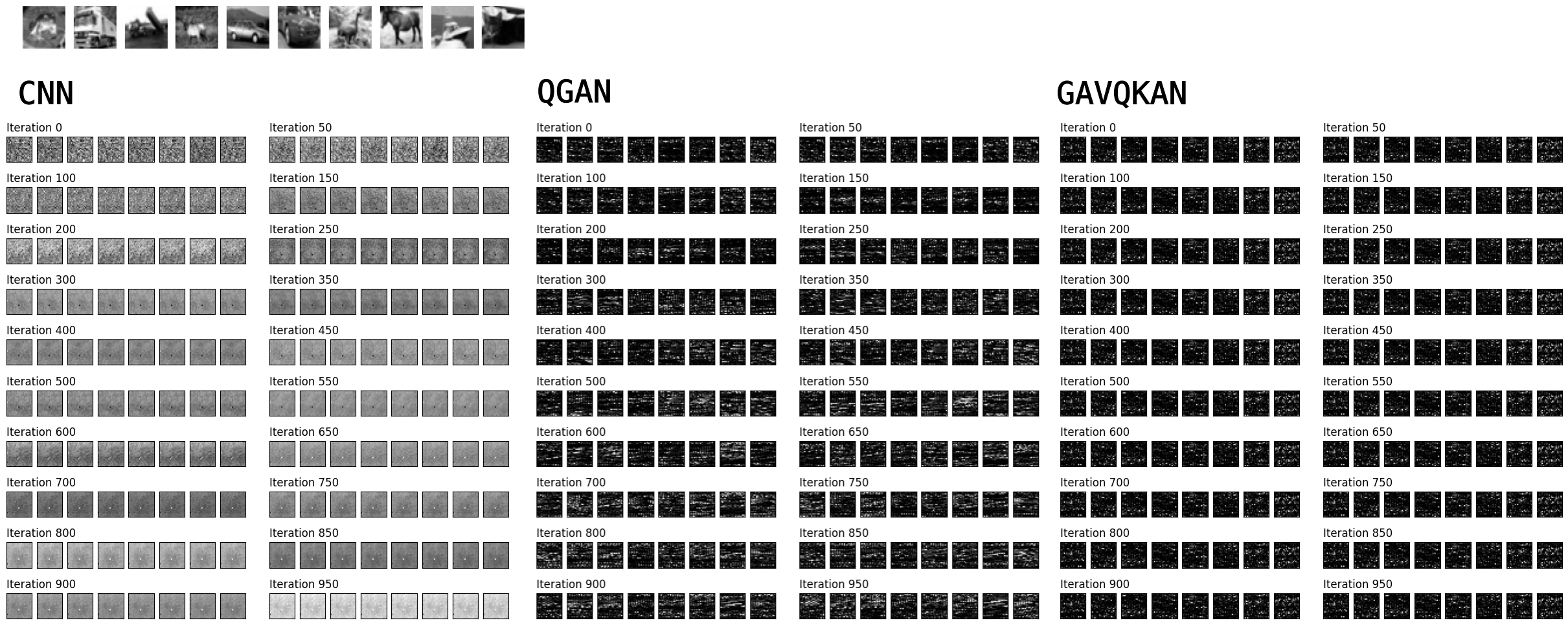}
\caption{ 
The generated picture of the first 8 data by GAVQKAN, QGAN, and CNN using the cross-entropy as a loss function.
The pictures in the upper right are the original pictures from 0 to 9 on the index. }\label{ gen a }
\end{figure}

\begin{figure}   

\includegraphics[scale=0.3]{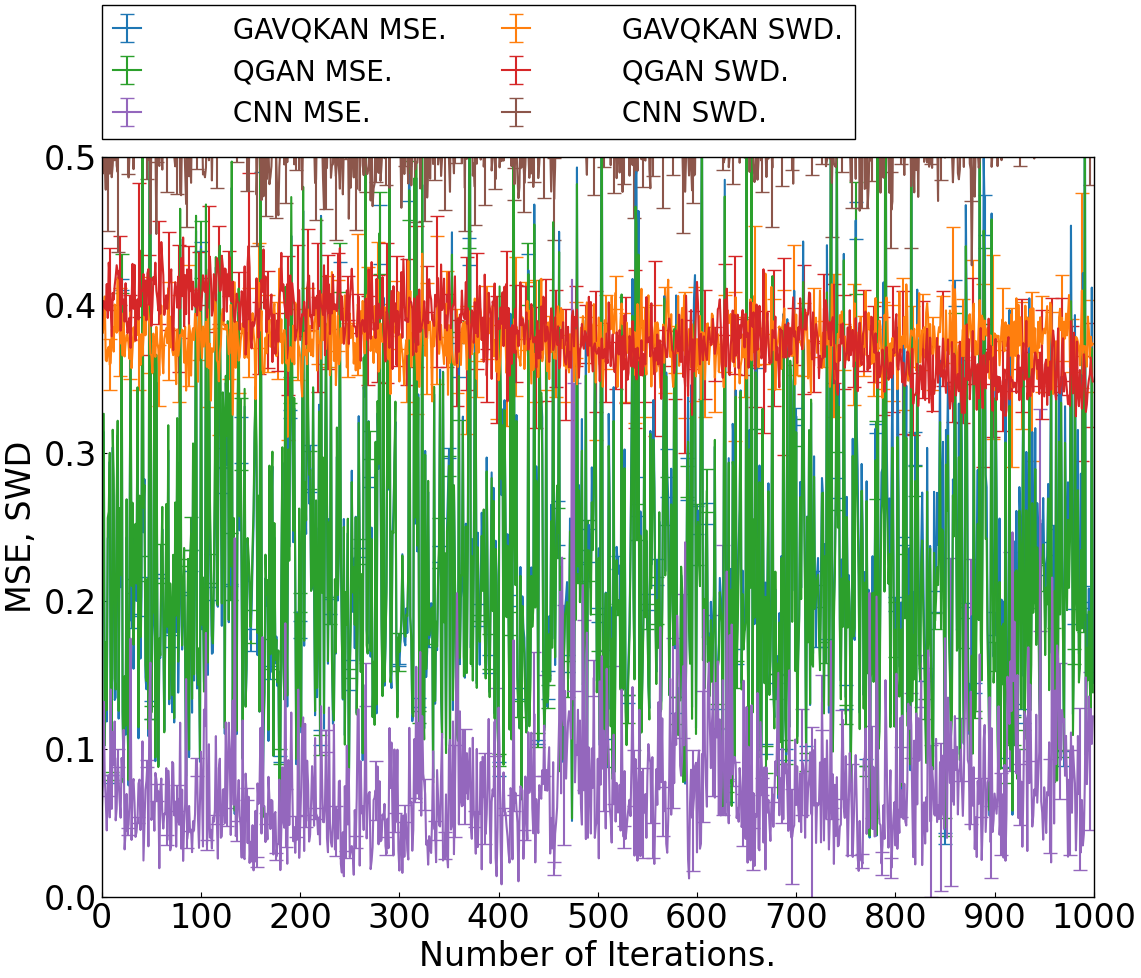}

\caption{ The number of iterations v.s. the average of the MSE and SWD for 5 attempts of GAVQKAN, QGAN, and CNN using cross-entropy as a loss function on the training and generating the CIFAR-10 dataset. } \label{ w d } 
\end{figure}

\begin{figure}     
    
\includegraphics[scale=0.3]{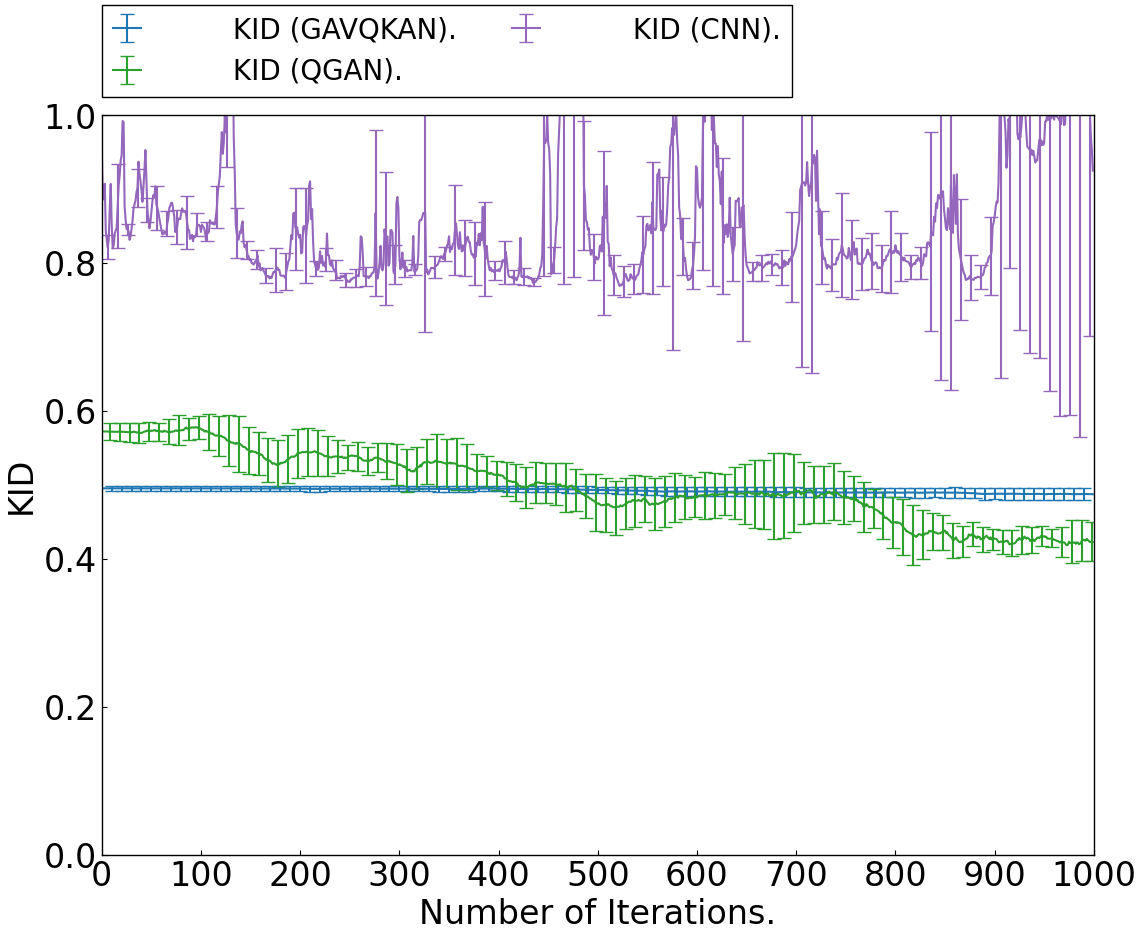}  
 
\caption{ The number of iterations v.s. the average of the KID for 5 attempts of GAVQKAN, QGAN, and CNN using cross-entropy as a loss function on the training and generating the CIFAR-10 dataset.  } \label{ w a }     
  
\end{figure}

Next, we compared the result of generating a 22 $ \times $ 22 size picture on the CIFAR-10 dataset.   
We used only 1000 data from index 0. 
Both pictures are compressed from 32 $ \times $ 32 size original pictures into 22 $ \times $ 22 size and converted to gray scale.
The number of qubits on QGAN is 8, and the size of the network on CNN is 8 times that of training on 0. 
We show the average of the loss function for 10 attempts of the discriminator and generator of GAVQKAN, QGAN, and CNN for the number of epochs using cross-entropy in Fig.\ref{ pic s }, and the generated picture of the first 8 pictures for fixed noise in Fig.\ref{ gen a }, respectively.
The average of GAVQKAN did not grow compared to QGAN and other cases of training, even though generated pictures have the smallest SWD in all, also shown in Fig.\ref{ w d }.
We show the MSE and SWD of three methods in Fig.\ref{ w d } and KID in Fig.\ref{ w a }, respectively.   
The loss functions of the generator of GAVQKAN became the largest in all, and that of QGAN is the second largest, the same as MNIST.        
In contrast, the loss functions of generators of GAVQKAN became the smallest, and suppressed the slowest in all.     
SWD of GAVQKAN in range the number of iterations below 400 is the smallest in all, and the second smallest over 400, the same as MNIST.  
MSE of GAVQKAN is on each learned data is the largest in all on all range.
KID of GAVQKAN is the smallest in the range; the number of iterations below 400 is the smallest, the same as SWD.     
It lowers gradually in all ranges different for others because the KID of each attempt of other methods saturates in the range over 900.    
Besides, the time for calculations is 4.8302 times larger than QGAN, 12.1289 times larger  than CNN, and  0.7437 times larger than Quantum Reservoir GAN ( QRGAN ) \cite{2025arXiv250805716W}, respectively.     
GAVQKAN takes less time than QRGAN because GAVQKAN emerges with longer data than MNIST.         
GAVQKAN exhibits high accuracy by small numbers of data as the case of MNIST.

    
\section{Discussion}\label{5}
\begin{figure}
\includegraphics[scale=0.3]{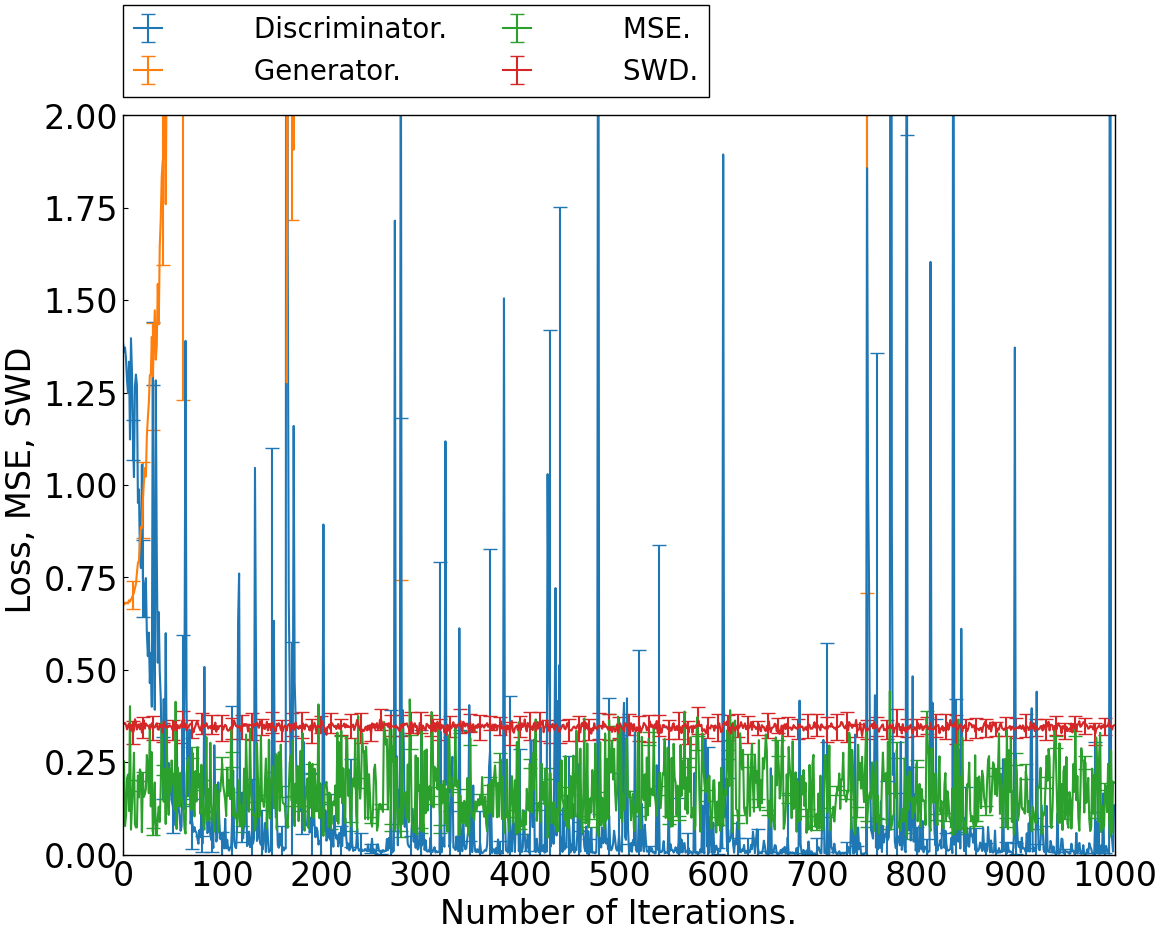}
\caption{The number of iterations v.s. the average of the loss function, MSE, and SWD for 16 seeds of the discriminator and generator of GAVQKAN using cross-entropy as a loss function.} \label{ pic f s }
\end{figure}

\begin{figure}
\includegraphics[scale=0.3]{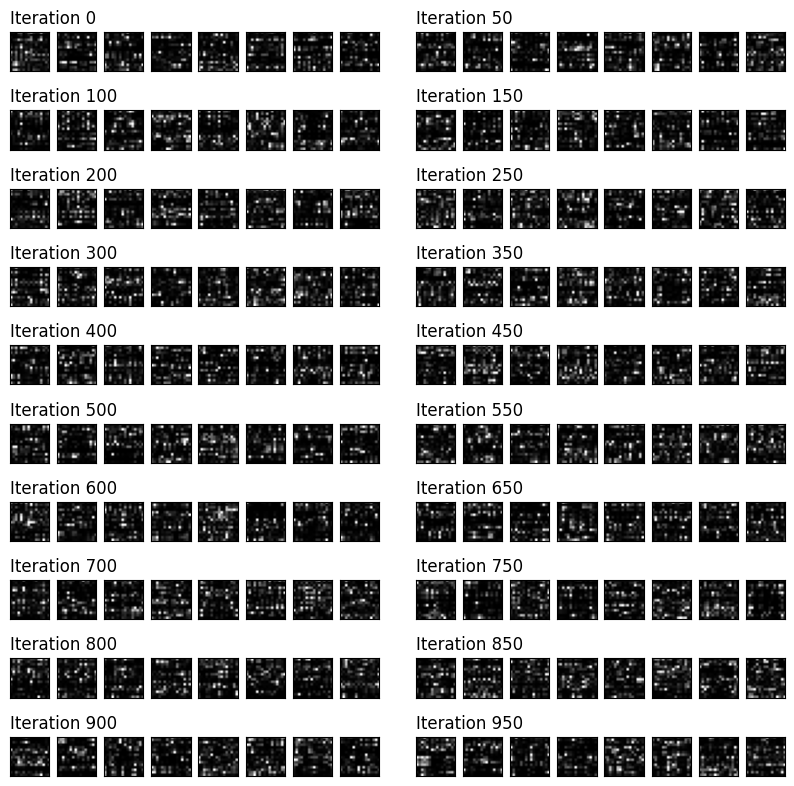}
\caption{The generated picture of the first 8 data for seed 0 by GAVQKAN using the cross-entropy as a loss function. }\label{ gen f a }
\end{figure}

In this section, we discuss the reproduction property and the accuracy for the noisy input of GAVQKAN. 
First, we discuss the reproduction property of the output of GAVQKAN for data on Fashion-MNIST.
We generated the first 8 data learning 1000 data using GAVQKAN, and calculated the loss functions of discriminator, the loss functions of generator, MSE and SWD for seed coefficient of random numbers $ 42, 2, 10, 13, 0, 3407, 7120, 10000, 11111, 16384, 17171, 130000, 14480, 11668, 500001 $ and $ 620000 $ and calculated t and p values by Bonferroni test.
Data are all original size ( $ 16 \times 16 $ ).
Then, we use radial base function instead of B spline function in calculation of $ \phi $s for saving calculation time. 
We show the average of the loss function, MSE, and SWD for 16 seeds of GAVQKAN for the number of epochs using cross-entropy in Fig.\ref{ pic f s }, and the generated picture of the first 8 pictures for fixed noise in Fig.\ref{ gen f a }, respectively.
The generator learned and was enlarged by the progress of learning.
The average of SWD is nearly 0.35, the same as the result of CIFAR-10.
According to the result of the Bonferroni test on SWD for all iterations, at least 9 seeds are clearly different from the given seed ( shown in Table.\ref { b t } ).
The significance is $ 0.05 / 240 $, hence, the two seeds that have p value over 0 differ negligibly.
SWD for different seeds has a little difference compared to QRGAN.
\begin{table*}[h]
\caption{ The value of t and p values between two seeds on Fashion-MNIST of SWD of GAVQKAN using cross-entropy as a loss function. }\label{ b t }
\centering
\resizebox{\textwidth}{!}{%
\begin{tabular}{c|c|c|c|c|c|c|c|c|c|c|c|c|c|c|c|c} \hline \hline
t value&42&2&10&13&0&3407&7120&10000&11111&16384&17171&130000&14480&11668&500001&620000\\\hline  
42&0&0.1605&4.2064&-0.6913&4.551&-0.0434&2.1098&0.2537&-1.2875&1.1109&0.789&1.7223&-0.631&4.2608&4.4995&-0.0683\\\hline  
2&-0.1605&0&3.9293&-0.8319&4.26&-0.2006&1.8909&0.0854&-1.4117&0.9199&0.6094&1.5178&-0.7715&3.9721&4.2177&-0.2251\\\hline  
10&-4.2064&-3.9293&0&-4.8694&0.2988&-4.1911&-2.0958&-3.974&-5.4579&-3.09&-3.3596&-2.4206&-4.7491&-0.0576&0.3315&-4.2253\\\hline  
13&0.6913&0.8319&4.8694&0&5.2185&0.6387&2.7868&0.9467&-0.5935&1.7939&1.4659&2.3921&0.0501&4.94&5.1549&0.6155\\\hline  
0&-4.551&-4.26&-0.2988&-5.2185&0&-4.5299&-2.4174&-4.3181&-5.8125&-3.422&-3.6898&-2.7403&-5.091&-0.3653&0.039&-4.5654\\\hline  
3407&0.0434&0.2006&4.1911&-0.6387&4.5299&0&2.1237&0.2937&-1.2268&1.1388&0.8211&1.7417&-0.58&4.2427&4.4807&-0.0245\\\hline  
7120&-2.1098&-1.8909&2.0958&-2.7868&2.4174&-2.1237&0&-1.8668&-3.379&-0.9963&-1.2919&-0.356&-2.6964&2.0941&2.4081&-2.1533\\\hline  
10000&-0.2537&-0.0854&3.974&-0.9467&4.3181&-0.2937&1.8668&0&-1.5454&0.8631&0.5427&1.4809&-0.8824&4.0231&4.2705&-0.3193\\\hline  
11111&1.2875&1.4117&5.4579&0.5935&5.8125&1.2268&3.379&1.5454&0&2.3873&2.0526&2.9765&0.6352&5.5435&5.7376&1.2049\\\hline  
16384&-1.1109&-0.9199&3.09&-1.7939&3.422&-1.1388&0.9963&-0.8631&-2.3873&0&-0.3077&0.6259&-1.7179&3.1146&3.3929&-1.1661\\\hline  
17171&-0.789&-0.6094&3.3596&-1.4659&3.6898&-0.8211&1.2919&-0.5427&-2.0526&0.3077&0&0.9219&-1.3956&3.3898&3.6567&-0.8474\\\hline  
130000&-1.7223&-1.5178&2.4206&-2.3921&2.7403&-1.7417&0.356&-1.4809&-2.9765&-0.6259&-0.9219&0&-2.3091&2.4264&2.7256&-1.77\\\hline  
14480&0.631&0.7715&4.7491&-0.0501&5.091&0.58&2.6964&0.8824&-0.6352&1.7179&1.3956&2.3091&0&4.8137&5.0322&0.557\\\hline  
11668&-4.2608&-3.9721&0.0576&-4.94&0.3653&-4.2427&-2.0941&-4.0231&-5.5435&-3.1146&-3.3898&-2.4264&-4.8137&0&0.3973&-4.2782\\\hline  
500001&-4.4995&-4.2177&-0.3315&-5.1549&-0.039&-4.4807&-2.4081&-4.2705&-5.7376&-3.3929&-3.6567&-2.7256&-5.0322&-0.3973&0&-4.5152\\\hline  
620000&0.0683&0.2251&4.2253&-0.6155&4.5654&0.0245&2.1533&0.3193&-1.2049&1.1661&0.8474&1.77&-0.557&4.2782&4.5152&0\\\hline   \hline

p value&42&2&10&13&0&3407&7120&10000&11111&16384&17171&130000&14480&11668&500001&620000\\\hline  
42&1&0.8725&0&0.4894&0&0.9654&0.035&0.7998&0.1981&0.2667&0.4302&0.0852&0.5281&0&0&0.9456\\\hline  
2&0.8725&1&0.0001&0.4056&0&0.841&0.0588&0.932&0.1582&0.3577&0.5423&0.1292&0.4405&0.0001&0&0.8219\\\hline  
10&0&0.0001&1&0&0.7651&0&0.0362&0.0001&0&0.002&0.0008&0.0156&0&0.9541&0.7403&0\\\hline  
13&0.4894&0.4056&0&1&0&0.5231&0.0054&0.3439&0.5529&0.073&0.1428&0.0168&0.96&0&0&0.5383\\\hline  
0&0&0&0.7651&0&1&0&0.0157&0&0&0.0006&0.0002&0.0062&0&0.715&0.9689&0\\\hline   
3407&0.9654&0.841&0&0.5231&0&1&0.0338&0.769&0.22&0.2549&0.4117&0.0817&0.562&0&0&0.9805\\\hline  
7120&0.035&0.0588&0.0362&0.0054&0.0157&0.0338&1&0.0621&0.0007&0.3192&0.1965&0.7219&0.0071&0.0364&0.0161&0.0314\\\hline  
10000&0.7998&0.932&0.0001&0.3439&0&0.769&0.0621&1&0.1224&0.3882&0.5874&0.1388&0.3777&0.0001&0&0.7495\\\hline  
11111&0.1981&0.1582&0&0.5529&0&0.22&0.0007&0.1224&1&0.0171&0.0402&0.003&0.5254&0&0&0.2284\\\hline  
16384&0.2667&0.3577&0.002&0.073&0.0006&0.2549&0.3192&0.3882&0.0171&1&0.7584&0.5314&0.086&0.0019&0.0007&0.2437\\\hline   
17171&0.4302&0.5423&0.0008&0.1428&0.0002&0.4117&0.1965&0.5874&0.0402&0.7584&1&0.3567&0.163&0.0007&0.0003&0.3969\\\hline  
130000&0.0852&0.1292&0.0156&0.0168&0.0062&0.0817&0.7219&0.1388&0.003&0.5314&0.3567&1&0.021&0.0153&0.0065&0.0769\\\hline  
14480&0.5281&0.4405&0&0.96&0&0.562&0.0071&0.3777&0.5254&0.086&0.163&0.021&1&0&0&0.5776\\\hline  
11668&0&0.0001&0.9541&0&0.715&0&0.0364&0.0001&0&0.0019&0.0007&0.0153&0&1&0.6912&0\\\hline  
500001&0&0&0.7403&0&0.9689&0&0.0161&0&0&0.0007&0.0003&0.0065&0&0.6912&1&0\\\hline  
620000&0.9456&0.8219&0&0.5383&0&0.9805&0.0314&0.7495&0.2284&0.2437&0.3969&0.0769&0.5776&0&0&1\\\hline  
\end{tabular}
}%
\end{table*}

Second, we discuss the dependency on depth $ N _ d $ of the output of GAVQKAN for data on MNIST.
We generated the first 8 data learning 1000 data using GAVQKAN and QGAN, and calculated the SWD and KID for depth $ N _ d = 1 $ to $ 8 $ comparing them.  
Data are all original size ( $ 16 \times 16 $ ).
Then, we use radial base function instead of B spline function in calculation of eq.\ref { k a n } for saving calculation time.

\begin{figure}

\includegraphics[scale= 0.2 ]{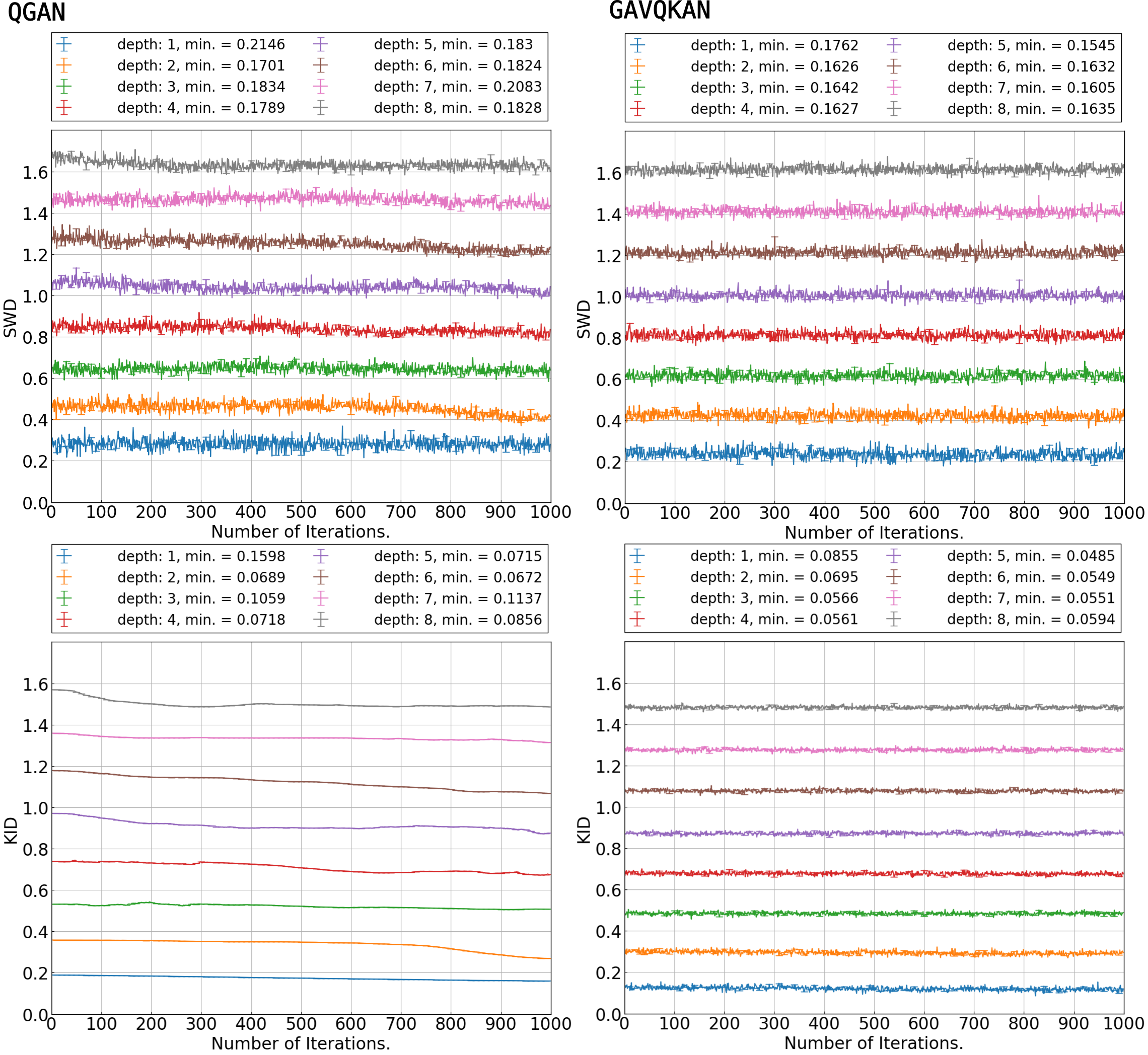}

\caption{The number of iterations v.s. the average of the SWD and KID of GAVQKAN and QGAN on the MNIST dataset using cross-entropy as a loss function.
Each graph of depth is 0.2 separated from neighbor graph of depth. } \label{ d c }

\end{figure}

\begin{figure}
 
\includegraphics[scale=0.3]{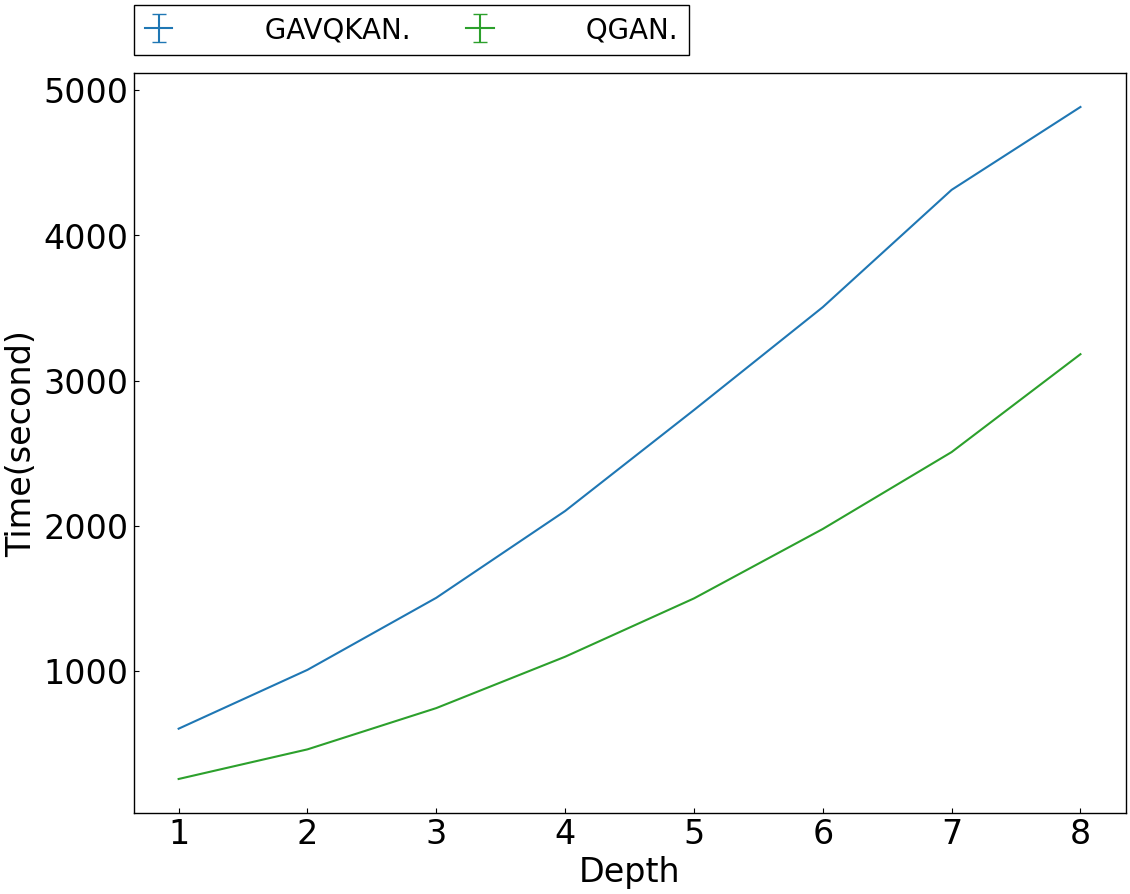}
 
\caption{The number of iterations v.s. the average of calculation time of GAVQKAN and QGAN on the MNIST dataset using cross-entropy as a loss function.} \label{ d t }

\end{figure}

We show the calculated the SWD and KID of GAVQKAN and QGAN on average of 5 attempts on Fig. \ref { d c }, and time for calculation on Fig. \ref { d t }, respectively.  
GAVQKAN exhibits higher accuracy than QGAN for less than 300 trained data reguard less of the depth with respect to the SWD and KID. 
Besides, GAVQKAN is more accurate than QGAN except in case the depth is 2.  
GAVQKAN is the most accurate when the depth is 5, and QGAN is the most accurate when the depth is 2, respectively.  
The time for calculation of both are parabolic for depth.  

\begin{figure}
\includegraphics[scale=0.3]{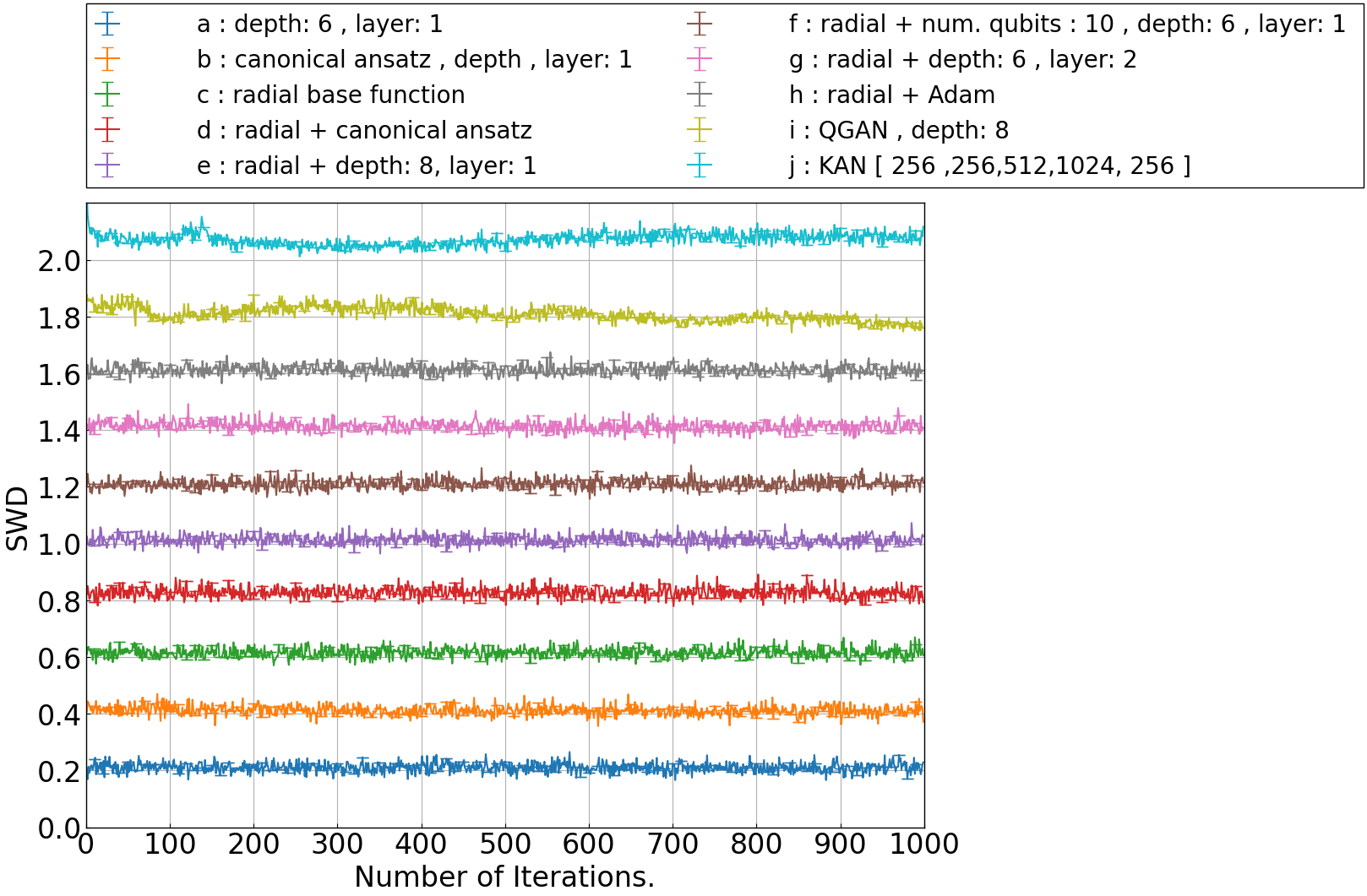}
\caption{The number of iterations v.s. the average of the SWD of  10 cases on the MNIST dataset using cross-entropy as a loss function.
Each graph of depth is 0.2 separated from neighbor graph of case.  } \label{ sep m }
\end{figure}

Next, we surveyed the accuracy of GAVQKAN for the case a : depth: 6, layer: 1, b : canonical ansatz, depth, layer: 1, c : radial base function, d : radial + canonical ansatz, e : radial + depth: 8, layer: 1, f : radial + num. qubits: 10, depth: 6, layer: 1, g : radial + depth: 6, layer: 2, h : radial + Adam, i : QGAN, depth: 8, j : KAN [ 256,256,512,1024,256 ], and discussed  the method to improve it on the MNIST dataset.  
The canonical ansatz is the ansatz of paper \cite{Wakaura_VQKAN_2024}, rbf indicates the case uses the radial base function instead of B spline function on eq.\ref { k a n }, and Adam indicates the optimizer is the Adam method.    
The canonical ansatz saves time for calculation and improves the accuracy a little compared to our ansatz, as shown in Fig. \ref { sep m } and Table. \ref { time }.
The radial base function instead of B spline function on KAN layers saved prominent time for calculation, prominently in exchange for only about 0.02 SWD on average, Fig. \ref { sep m } and Table. \ref { time }.  
Cases e and f improved the accuracy and reached the accuracy of QGAN including case i.  
Adam optimizer saves time for calculation further.      
However, the loss functions converged far faster than all other cases.
Case a-h are all lowered  a little or nearly never changed differently from cases i and j. 
We suppose that the learning rates of the discriminator and generator  are inadequate to train them enough.  
Tuning them may be read to improve accuracy. 
Besides, applying APA(Adaptive Pseudo Augmentation) \cite{2021arXiv211106849J} and other improved GAN \cite{9770060,2021arXiv211101118K} can improve the accuracy.  
 
Improvement of the method of GAN also improve the accuracy. 

Degradation Guided Generative Adversarial Network \cite{2025arXiv251207253X} shows the prominent improvement of accuracy and robustness against the data.  
  
This may supress the spreading of effect of quantum noise.  
 
Other small improvement of loss function such as Least-Squares GAN \cite{2016arXiv161104076M} may also improve the accuracy \cite{2017arXiv170400028G}.

KAN itself requires a large number of iterations for convergence.
The robustness of VQKAN against overfitting may lower the speed of convergence.

\begin{table}[h]
\caption{ The speed of calculations on each case for 1000 iterations. The time for calculation of case a is $ t _ 0 $.  }\label{ time }

\begin{tabular}{c|c} \hline \hline  
case&speed ( $ t _ 0 / t _ n $ )\\\hline   
a&1\\\hline     
b &1.3134\\\hline   
c &3.6527\\\hline     
d &3.6966\\\hline   
e &2.6299\\\hline    
f&3.1761\\\hline    
g &1.8904\\\hline    
h&9.4137\\\hline  
i&4.7942\\\hline  
j&51.749\\\hline  
\end{tabular}
 
\end{table}

\begin{figure}    

\includegraphics[scale=0.3]{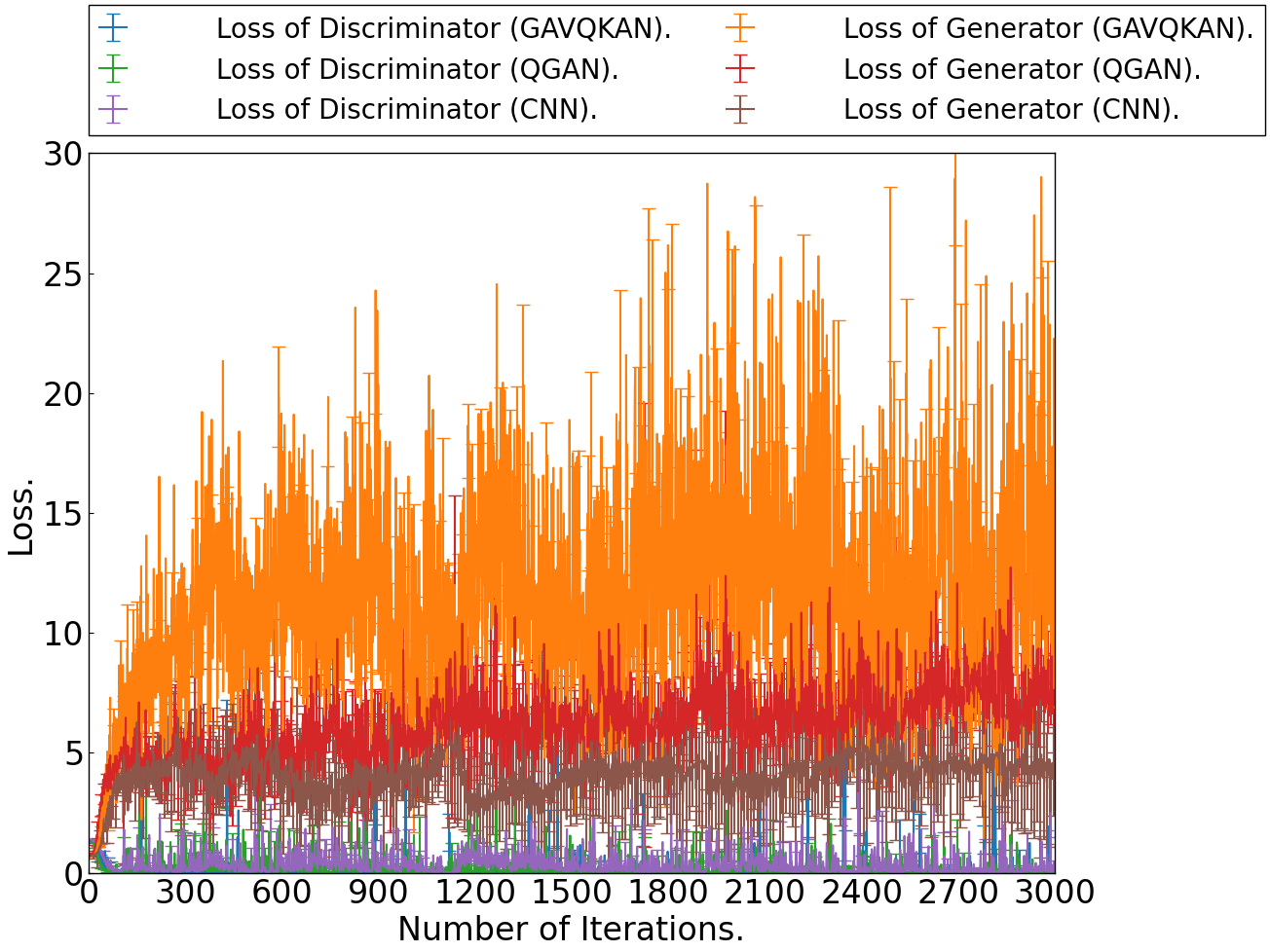}

\caption{ The number of iterations v.s. the average of the loss function for 10 attempts of the discriminator $ L_D$ and generator $ l_G$ of GAVQKAN, QGAN, and CNN using  cross-entropy as a loss function 8 $ \times $  8-sized MNIST for 3000 samples.} \label{ comp 0 n l }

\end{figure}

\begin{figure}   

\includegraphics[scale=0.3]{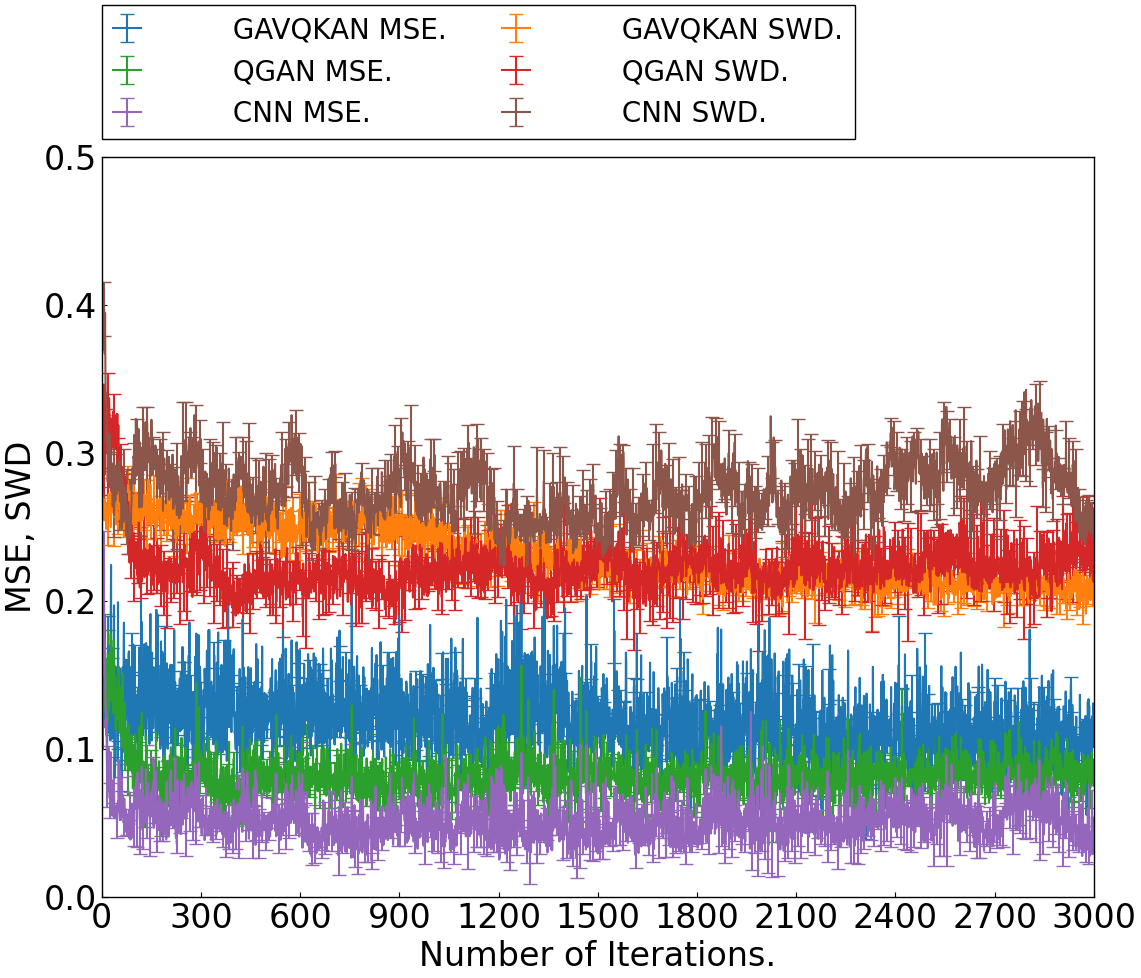}

\caption{ The number of iterations v.s. the average of the MSE and SWD for 5 attempts of GAVQKAN, QGAN, and CNN using cross-entropy as a loss function on the training and generating the 8 $ \times $  8-sized MNIST dataset. } \label{ s l }

\end{figure}

\begin{figure}  
   
\includegraphics[scale=0.3]{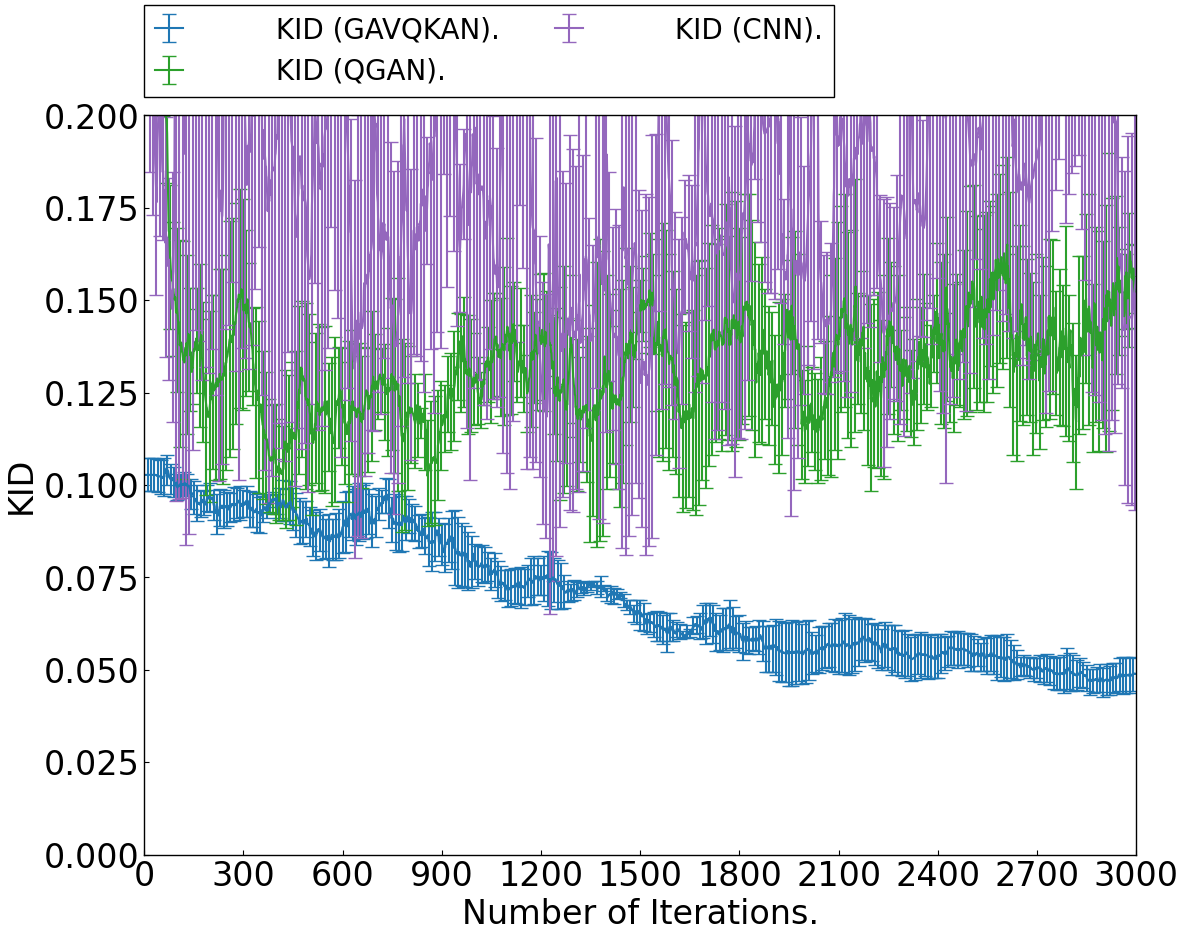}  

\caption{ The number of iterations v.s. the average of the KID for 5 attempts of GAVQKAN, QGAN, and CNN using cross-entropy as a loss function on the training and generating the 8 $ \times $  8-sized MNIST dataset. } \label{ d l }

\end{figure}

We also surveyed whether the training of GAVQKAN saturates or not for the number of iterations.  
We compared the results on GAVQKAN, QGAN, and Classical Neural Network (CNN) using loss functions of cross-entropy on 8 $ \times $  8-sized MNIST for 3000 samples, respectively.  
Then $ N _ q $ of GAVQKAN and QGAN are 5 including 1 ancilla qubit.  
We show the average of the loss function for 10 attempts of the discriminator and generator of GAVQKAN, QGAN, and CNN for the number of epochs using cross-entropy in Fig.\ref{ comp 0 n l }, the SWD and MSE in Fig.\ref { s l }, and the KID in Fig.\ref{ d l }, respectively.    
The loss function the same trends as 16 $ \times $ 16-sized case, and the loss function of generator of GAVQKAN became the largest among all.  
The SWD of GAVQKAN is the second smallest in the range over 300 and under 1500 for the the number of iterations on average.  
Although, it became the smallest decreasing gradually after that of QGAN saturated.  
The MSE of GAVQKAN is on each learned data is the largest in all on all range even the number of data is 3000, hence, GAVQKAN is supposed to be less trapped by local minimum than QGAN and CNN. 
The KID of GAVQKAN decreased monotonically in all range, and the smallest.  
It will saturate in range over 5000. 
These result have same tendency as the result of above sections in the range below 1000, hence, this tendency is universal at least on this setting for datasets.  
The GAVQKAN can learn at least 3 times larger number of data than QGAN even if the ansatz is same, hence, it has the potential to be more accurate.  
Even the convergence of GAVQKAN is slower than other  methods \cite{ma_quantum_2025,math12233852}, GAVQKAN has larger accomodation for data \cite{Wakaura_VQKAN_2024}.

  \newpage
  
  \clearpage

\section{Concluding remarks}\label{7}
In this paper, we revealed that GAVQKAN has more accuracy than QGAN, CNN, and classical KAN on the MNIST and CIFAR-10 datasets for small amounts of data.  
Especially, GAVQKAN exhibits the highest accuracy than QGAN, CNN, and classical KAN on the MNIST for large number of data without leaching pleateau.  
This result may contribute to the generation of large data by a larger number of qubits. 
Besides, the accuracy for the number of parameters is smaller than classical KAN.
However, the speed of convergence is slower than that of CNN and classical KAN. 
Improving it is the next problem.
Our work is only on simulators; hence, calculating on real quantum devices is also the next plan.

In addition, our method of applying KAN is applicable for other GAN such as Quantum Implicit Neural Representation based QGAN \cite{ma_quantum_2025} and quantum-classical architecture for improving GAN \cite{2024arXiv240201791S}, hence, they will have synergy.  

\section*{ Data availability }

The data that support the findings of this study are available from the corresponding author, Hikaru Wakaura, upon reasonable request.

\section*{ Author Declarations  }

\subsection* { Conflict of Interest  } 

The authors have no conflicts to disclose.  
 
\subsection* {  Author Contributions   }   

Hikaru Wakaura : Conceptualization (lead); Data curation (lead); Formal analysis (lead); Investigation (lead); Methodology (lead); Project administration (lead); Visualization (lead); Writing – original draft (lead); Writing – review$ \& $editing (lead).

\bibliography{mainwakaura2}

@Article{McClean_2016,
  author    = {Jarrod R McClean and Jonathan Romero and Ryan Babbush and Al{\'{a}}n Aspuru-Guzik},
  title     = {The theory of variational hybrid quantum-classical algorithms},
  journal   = {New Journal of Physics},
  year      = {2016},
  volume    = {18},
  number    = {2},
  pages     = {023023},
  month     = {feb},



doi       = {10.1088/1367-2630/18/2/023023},
  publisher = {{IOP} Publishing},
  url       = {https://doi.org/10.1088%2F1367-2630%2F18%2F2%2F023023},
}

@Article{PhysRevA.98.032309,
  author    = {Mitarai, K. and Negoro, M. and Kitagawa, M. and Fujii, K.},
  title     = {Quantum circuit learning},
  journal   = {Phys. Rev. A},
  year      = {2018},
  volume    = {98},
  pages     = {032309},
  month     = {Sep},
  doi       = {10.1103/PhysRevA.98.032309},
  issue     = {3},
  numpages  = {6},
  publisher = {American Physical Society},
  url       = {https://link.aps.org/doi/10.1103/PhysRevA.98.032309},
}

@Article{2019arXiv190608728P,
  author        = {{Parrish}, Robert M. and {Hohenstein}, Edward G. and {McMahon}, Peter L. and {Martinez}, Todd J.},
  title         = {{Hybrid Quantum/Classical Derivative Theory: Analytical Gradients and Excited-State Dynamics for the Multistate Contracted Variational Quantum Eigensolver}},
  journal       = {arXiv e-prints},
  year          = {2019},
  pages         = {arXiv:1906.08728},
  month         = jun,
  adsurl        = {https://ui.adsabs.harvard.edu/abs/2019arXiv190608728P},
  archiveprefix = {arXiv},
  eid           = {arXiv:1906.08728},
  eprint        = {1906.08728},
  primaryclass  = {quant-ph},
}

@Article{Grimsley2019,
  author        = {{Grimsley}, Harper R. and {Economou}, Sophia E. and {Barnes}, Edwin and {Mayhall}, Nicholas J.},
  title         = {{An adaptive variational algorithm for exact molecular simulations on a quantum computer}},
  journal       = {Nature Communications},
  year          = {2019},
  volume        = {10},
  pages         = {3007},
  adsurl        = {https://ui.adsabs.harvard.edu/abs/2019NatCo..10.3007G},
  archiveprefix = {arXiv},
  doi           = {10.1038/s41467-019-10988-2},
  eid           = {3007},
  eprint        = {1812.11173},
  primaryclass  = {quant-ph},
}

@Article{feynman_simulating_1982,
  author  = {Feynman, Richard P.},
  title   = {Simulating physics with computers},
  journal = {International Journal of Theoretical Physics},
  year    = {1982},
  volume  = {21},
  number  = {6},
  pages   = {467--488},
  month   = jun,
  issn    = {1572-9575},
  doi     = {10.1007/BF02650179},
  url     = {https://doi.org/10.1007/BF02650179},
}

@Article{2019QS&T....4a4001K,
  author        = {{Khoshaman}, Amir and {Vinci}, Walter and {Denis}, Brandon and {Andriyash}, Evgeny and {Sadeghi}, Hossein and {Amin}, Mohammad H.},
  title         = {{Quantum variational autoencoder}},
  journal       = {Quantum Science and Technology},
  year          = {2019},
  volume        = {4},
  number        = {1},
  pages         = {014001},
  month         = jan,
  adsurl        = {https://ui.adsabs.harvard.edu/abs/2019QS&T....4a4001K},
  archiveprefix = {arXiv},
  doi           = {10.1088/2058-9565/aada1f},
  eprint        = {1802.05779},
  primaryclass  = {quant-ph},
}

@Article{2019Natur.567..209H,
  author        = {{Havl{\'\i}{\v{c}}ek}, Vojt{\v{e}}ch and {C{\'o}rcoles}, Antonio D. and {Temme}, Kristan and {Harrow}, Aram W. and {Kandala}, Abhinav and {Chow}, Jerry M. and {Gambetta}, Jay M.},
  title         = {{Supervised learning with quantum-enhanced feature spaces}},
  journal       = {Nature},
  year          = {2019},
  volume        = {567},
  number        = {7747},
  pages         = {209-212},
  month         = mar,
  adsurl        = {https://ui.adsabs.harvard.edu/abs/2019Natur.567..209H},
  archiveprefix = {arXiv},
  doi           = {10.1038/s41586-019-0980-2},
  eprint        = {1804.11326},
  primaryclass  = {quant-ph},
}

@Article{2021PhRvP..16d4057B, 
  author        = {{Benedetti}, Marcello and {Coyle}, Brian and {Fiorentini}, Mattia and {Lubasch}, Michael and {Rosenkranz}, Matthias},
  title         = {{Variational Inference with a Quantum Computer}},
  journal       = {Physical Review Applied},
  year          = {2021},
  volume        = {16},
  number        = {4},
  pages         = {044057},
  month         = oct,
  adsurl        = {https://ui.adsabs.harvard.edu/abs/2021PhRvP..16d4057B},
  archiveprefix = {arXiv},
  doi           = {10.1103/PhysRevApplied.16.044057},
  eid           = {044057},
  eprint        = {2103.06720},
  primaryclass  = {quant-ph},
}

@Article{2022PhRvA.106b2601A,
  author        = {{Abel}, Steve and {Criado}, Juan C. and {Spannowsky}, Michael},
  title         = {{Completely quantum neural networks}},
  journal       = {Phys. Rev. A},
  year          = {2022},
  volume        = {106},
  number        = {2},
  pages         = {022601},
  month         = aug,
  adsurl        = {https://ui.adsabs.harvard.edu/abs/2022PhRvA.106b2601A},
  archiveprefix = {arXiv},
  doi           = {10.1103/PhysRevA.106.022601},
  eid           = {022601},
  eprint        = {2202.11727},
  primaryclass  = {quant-ph},
}

@Article{2022arXiv220211200K,
  author        = {{Kwak}, Yunseok and {Yun}, Won Joon and {Pyoung Kim}, Jae and {Cho}, Hyunhee and {Choi}, Minseok and {Jung}, Soyi and {Kim}, Joongheon},
  title         = {{Quantum Distributed Deep Learning Architectures: Models, Discussions, and Applications}},
  journal       = {arXiv e-prints},
  year          = {2022},
  pages         = {arXiv:2202.11200},
  month         = feb,
  adsurl        = {https://ui.adsabs.harvard.edu/abs/2022arXiv220211200K},
  archiveprefix = {arXiv},
  eid           = {arXiv:2202.11200},
  eprint        = {2202.11200},
  primaryclass  = {quant-ph},
}

@Article{2020PhRvL.125j0401W,
  author        = {{Wang}, Zhikang T. and {Ashida}, Yuto and {Ueda}, Masahito},
  title         = {{Deep Reinforcement Learning Control of Quantum Cartpoles}},
  journal       = {Physical Review Letters},
  year          = {2020},
  volume        = {125},
  number        = {10},
  pages         = {100401},
  month         = sep,
  adsurl        = {https://ui.adsabs.harvard.edu/abs/2020PhRvL.125j0401W},
  archiveprefix = {arXiv},
  doi           = {10.1103/PhysRevLett.125.100401},
  eid           = {100401},
  eprint        = {1910.09200},
  primaryclass  = {quant-ph},
}

@Article{2014PhRvL.113m0503R,
  author        = {{Rebentrost}, Patrick and {Mohseni}, Masoud and {Lloyd}, Seth},
  title         = {{Quantum Support Vector Machine for Big Data Classification}},
  journal       = {Phys. Rev. Lett.},
  year          = {2014},
  volume        = {113},
  number        = {13},
  pages         = {130503},
  month         = sep,
  adsurl        = {https://ui.adsabs.harvard.edu/abs/2014PhRvL.113m0503R},
  archiveprefix = {arXiv},
  doi           = {10.1103/PhysRevLett.113.130503},
  eid           = {130503},
  eprint        = {1307.0471},
  primaryclass  = {quant-ph},
}

@Article{2024EPJQT..11...76K,
  author        = {{Kundu}, Akash and {Sarkar}, Aritra and {Sadhu}, Abhishek},
  title         = {{KANQAS: Kolmogorov-Arnold Network for Quantum Architecture Search}},
  journal       = {EPJ Quantum Technology},
  year          = {2024},
  volume        = {11},
  number        = {1},
  pages         = {76},
  month         = dec,
  adsurl        = {https://ui.adsabs.harvard.edu/abs/2024EPJQT..11...76K},
  archiveprefix = {arXiv},
  doi           = {10.1140/epjqt/s40507-024-00289-z},
  eid           = {76},
  eprint        = {2406.17630},
  primaryclass  = {quant-ph},
}

@Article{2024arXiv241004435I, 
  author        = {{Ivashkov}, Petr and {Huang}, Po-Wei and {Koor}, Kelvin and {Pira}, Lirand{\"e} and {Rebentrost}, Patrick}, 
  title         = {{QKAN: Quantum Kolmogorov-Arnold Networks}},
  journal       = {arXiv e-prints},
  year          = {2024},
  pages         = {arXiv:2410.04435},
  month         = oct,
  adsurl        = {https://ui.adsabs.harvard.edu/abs/2024arXiv241004435I},
  archiveprefix = {arXiv},
  doi           = {10.48550/arXiv.2410.04435},
  eid           = {arXiv:2410.04435},
  eprint        = {2410.04435},
  primaryclass  = {quant-ph},
}

@Article{2024arXiv241007446J,
  author        = {{Jahin}, Md Abrar and {Akmol Masud}, Md. and {Mridha}, M.~F. and {Aung}, Zeyar and {Dey}, Nilanjan},
  title         = {{KACQ-DCNN: Uncertainty-Aware Interpretable Kolmogorov-Arnold Classical-Quantum Dual-Channel Neural Network for Heart Disease Detection}},
  journal       = {arXiv e-prints}, 
  year          = {2024}, 
  pages         = {arXiv:2410.07446}, 
  month         = oct,
  adsurl        = {https://ui.adsabs.harvard.edu/abs/2024arXiv241007446J},
  archiveprefix = {arXiv},
  doi           = {10.48550/arXiv.2410.07446},
  eid           = {arXiv:2410.07446},
  eprint        = {2410.07446},
  primaryclass  = {cs.LG},
}

@Article{2024arXiv241114902K,
  author        = {{Kou}, Wei and {Chen}, Xurong},
  title         = {{Machine Learning Insights into Quark-Antiquark Interactions: Probing Field Distributions and String Tension in QCD}},
  journal       = {arXiv e-prints},
  year          = {2024},
  pages         = {arXiv:2411.14902},
  month         = nov,
  adsurl        = {https://ui.adsabs.harvard.edu/abs/2024arXiv241114902K},
  archiveprefix = {arXiv},
  doi           = {10.48550/arXiv.2411.14902},
  eid           = {arXiv:2411.14902},
  eprint        = {2411.14902},
  primaryclass  = {hep-ph},
}

@Article{2024arXiv241203710K,
  author        = {{Kim}, Taehyeun and {Girard}, Anouck and {Kolmanovsky}, Ilya},
  title         = {{CIKAN: Constraint Informed Kolmogorov-Arnold Networks for Autonomous Spacecraft Rendezvous using Time Shift Governor}},
  journal       = {arXiv e-prints},
  year          = {2024},
  pages         = {arXiv:2412.03710},
  month         = dec,
  adsurl        = {https://ui.adsabs.harvard.edu/abs/2024arXiv241203710K},
  archiveprefix = {arXiv},
  doi           = {10.48550/arXiv.2412.03710},
  eid           = {arXiv:2412.03710},
  eprint        = {2412.03710},
  primaryclass  = {eess.SY},
}

@Article{2024arXiv241008452B,
  author        = {{Bandyopadhyay}, Yagnik and {Avlani}, Harshil and {Zhuang}, Houlong L.},
  title         = {{Kolmogorov-Arnold Neural Networks for High-Entropy Alloys Design}},
  journal       = {arXiv e-prints},
  year          = {2024},
  pages         = {arXiv:2410.08452},
  month         = oct,
  adsurl        = {https://ui.adsabs.harvard.edu/abs/2024arXiv241008452B},
  archiveprefix = {arXiv},
  doi           = {10.48550/arXiv.2410.08452},
  eid           = {arXiv:2410.08452},
  eprint        = {2410.08452},
  primaryclass  = {cond-mat.mtrl-sci},
}

@Article{2024arXiv241106727C,
  author        = {{Cang}, Yueyang and {liu}, Yu hang and {Shi}, Li},
  title         = {{Can KAN Work? Exploring the Potential of Kolmogorov-Arnold Networks in Computer Vision}},
  journal       = {arXiv e-prints},
  year          = {2024},
  pages         = {arXiv:2411.06727},
  month         = nov,
  adsurl        = {https://ui.adsabs.harvard.edu/abs/2024arXiv241106727C},
  archiveprefix = {arXiv},
  doi           = {10.48550/arXiv.2411.06727},
  eid           = {arXiv:2411.06727},
  eprint        = {2411.06727},
  primaryclass  = {cs.CV},
}

@ARTICLE{2024arXiv241118165H,
       author = {{Han}, Dong and {Li}, Yong and {Denzler}, Joachim},
        title = "{KAN See Your Face}",
      journal = {arXiv e-prints}, 
         year = 2024,
        month = nov,
          eid = {arXiv:2411.18165},
        pages = {arXiv:2411.18165}, 
          doi = {10.48550/arXiv.2411.18165},
archivePrefix = {arXiv},
       eprint = {2411.18165},
 primaryClass = {cs.CV}, 
       adsurl = {https://ui.adsabs.harvard.edu/abs/2024arXiv241118165H}
}

@ARTICLE{2024arXiv240800273T,
       author = {{Tang}, Tianze and {Chen}, Yanbing and {Shu}, Hai},
        title = "{3D U-KAN Implementation for Multi-modal MRI Brain Tumor Segmentation}",
      journal = {arXiv e-prints},
         year = 2024,
        month = aug,
          eid = {arXiv:2408.00273},
        pages = {arXiv:2408.00273},
          doi = {10.48550/arXiv.2408.00273},
archivePrefix = {arXiv},
       eprint = {2408.00273},
 primaryClass = {eess.IV},
       adsurl = {https://ui.adsabs.harvard.edu/abs/2024arXiv240800273T}
}

@Article{Wakaura_VQKAN_2024,
  author  = {{Wakaura}, Hikaru and {Bayu Suksmono}, Andriyan and {Mulyawan}, Rahmat},
  title   = {Variational Quantum Kolmogorov-Arnold Network},
  journal = {Research Square},
  year    = {2024},
  month   = July,
  note    = {PREPRINT (Version 3)},
  doi     = {10.21203/rs.3.rs-4504342/v3},
  url     = { https://doi.org/10.21203/rs.3.rs-4504342/v3   },
}

@Article{2021arXiv210501141W,
  author        = {{Wakaura}, Hikaru and {Bayu Suksmono}, Andriyan},
  title         = {{Tangent Vector Variational Quantum Eigensolver: A Robust Variational Quantum Eigensolver against the inaccuracy of derivative}},
  journal       = {arXiv e-prints},
  year          = {2021},
  pages         = {arXiv:2105.01141},
  month         = may,
  adsurl        = {https://ui.adsabs.harvard.edu/abs/2021arXiv210501141W},
  archiveprefix = {arXiv},
  doi           = {10.48550/arXiv.2105.01141},
  eid           = {arXiv:2105.01141},
  eprint        = {2105.01141},
  primaryclass  = {quant-ph},
}

@Article{2021arXiv210902009W,
  author        = {{Wakaura}, Hikaru and {Tomono}, Takao},
  title         = {{Genetic-Multi-initial Generalized VQE: Advanced VQE method using Genetic Algorithms then Local Search}},
  journal       = {arXiv e-prints},
  year          = {2021},
  pages         = {arXiv:2109.02009},
  month         = sep,
  adsurl        = {https://ui.adsabs.harvard.edu/abs/2021arXiv210902009W},
  archiveprefix = {arXiv},
  doi           = {10.48550/arXiv.2109.02009},
  eid           = {arXiv:2109.02009},
  eprint        = {2109.02009},
  primaryclass  = {quant-ph},
}

@Article{2025arXiv250805716W,
  author        = {{Wakaura}, Hikaru},
  title         = {{Quantum Reservoir GAN}},
  journal       = {arXiv e-prints},
  year          = {2025},
  pages         = {arXiv:2508.05716},
  month         = aug,
  adsurl        = {https://ui.adsabs.harvard.edu/abs/2025arXiv250805716W},
  archiveprefix = {arXiv},
  doi           = {10.48550/arXiv.2508.05716},
  eid           = {arXiv:2508.05716},
  eprint        = {2508.05716},
  primaryclass  = {quant-ph},
}

@Article{2021PhRvP..16b4051H,
  author        = {{Huang}, He-Liang and {Du}, Yuxuan and {Gong}, Ming and {Zhao}, Youwei and {Wu}, Yulin and {Wang}, Chaoyue and {Li}, Shaowei and {Liang}, Futian and {Lin}, Jin and {Xu}, Yu and {Yang}, Rui and {Liu}, Tongliang and {Hsieh}, Min-Hsiu and {Deng}, Hui and {Rong}, Hao and {Peng}, Cheng-Zhi and {Lu}, Chao-Yang and {Chen}, Yu-Ao and {Tao}, Dacheng and {Zhu}, Xiaobo and {Pan}, Jian-Wei},
  title         = {{Experimental Quantum Generative Adversarial Networks for Image Generation}},
  journal       = {Physical Review Applied},
  year          = {2021},
  volume        = {16},
  number        = {2},
  pages         = {024051},
  month         = aug,
  adsurl        = {https://ui.adsabs.harvard.edu/abs/2021PhRvP..16b4051H},
  archiveprefix = {arXiv},
  doi           = {10.1103/PhysRevApplied.16.024051},
  eid           = {024051},
  eprint        = {2010.06201},
  primaryclass  = {quant-ph},
}

@Article{math12233852,
  author         = {Pajuhanfard, Mohammadsaleh and Kiani, Rasoul and Sheng, Victor S.},
  title          = {Survey of Quantum Generative Adversarial Networks (QGAN) to Generate Images},
  journal        = {Mathematics},
  year           = {2024},
  volume         = {12},
  number         = {23},
  issn           = {2227-7390},
  article-number = {3852},
  doi            = {10.3390/math12233852},
  url            = {https://www.mdpi.com/2227-7390/12/23/3852},
}

@Article{JMLR:v22:20-451,
  author  = {Rémi Flamary and Nicolas Courty and Alexandre Gramfort and Mokhtar Z. Alaya and Aurélie Boisbunon and Stanislas Chambon and Laetitia Chapel and Adrien Corenflos and Kilian Fatras and Nemo Fournier and Léo Gautheron and Nathalie T.H. Gayraud and Hicham Janati and Alain Rakotomamonjy and Ievgen Redko and Antoine Rolet and Antony Schutz and Vivien Seguy and Danica J. Sutherland and Romain Tavenard and Alexander Tong and Titouan Vayer},
  title   = {POT: Python Optimal Transport},
  journal = {Journal of Machine Learning Research}, 
  year    = {2021},
  volume  = {22},
  number  = {78}, 
  pages   = {1--8},
  url     = {http://jmlr.org/papers/v22/20-451.html},
}

@Article{2023arXiv230609122S,
  author        = {{Singleton}, Jr, Robert L},
  title         = {{Shor's Factoring Algorithm and Modular Exponentiation Operators}},
  journal       = {arXiv e-prints},
  year          = {2023},
  pages         = {arXiv:2306.09122},
  month         = jun,
  adsurl        = {https://ui.adsabs.harvard.edu/abs/2023arXiv230609122S},
  archiveprefix = {arXiv},
  doi           = {10.48550/arXiv.2306.09122},
  eid           = {arXiv:2306.09122},
  eprint        = {2306.09122},
  primaryclass  = {quant-ph},
}

@Article{2023arXiv230504908M,
  author        = {{Mande}, Nikhil S. and {de Wolf}, Ronald},
  title         = {{Tight Bounds for Quantum Phase Estimation and Related Problems}},
  journal       = {arXiv e-prints},
  year          = {2023},
  pages         = {arXiv:2305.04908},
  month         = may,
  adsurl        = {https://ui.adsabs.harvard.edu/abs/2023arXiv230504908M},
  archiveprefix = {arXiv},
  doi           = {10.48550/arXiv.2305.04908},
  eid           = {arXiv:2305.04908}, 
  eprint        = {2305.04908},
  primaryclass  = {quant-ph},
}

@Article{PRXQuantum.4.040318,
  author    = {Chen, Jielun and Stoudenmire, E.M. and White, Steven R.},
  title     = {Quantum Fourier Transform Has Small Entanglement},
  journal   = {PRX Quantum},
  year      = {2023},
  volume    = {4},
  pages     = {040318},
  month     = {Oct},
  doi       = {10.1103/PRXQuantum.4.040318},
  issue     = {4},
  numpages  = {31},
  publisher = {American Physical Society},
  url       = {https://link.aps.org/doi/10.1103/PRXQuantum.4.040318},
}

@Article{2016arXiv161104076M,
  author        = {{Mao}, Xudong and {Li}, Qing and {Xie}, Haoran and {Lau}, Raymond Y.~K. and {Wang}, Zhen and {Smolley}, Stephen Paul},
  title         = {{Least Squares Generative Adversarial Networks}},
  journal       = {arXiv e-prints},
  year          = {2016},
  pages         = {arXiv:1611.04076},
  month         = nov,
  adsurl        = {https://ui.adsabs.harvard.edu/abs/2016arXiv161104076M},
  archiveprefix = {arXiv},
  doi           = {10.48550/arXiv.1611.04076},
  eid           = {arXiv:1611.04076},
  eprint        = {1611.04076},
  primaryclass  = {cs.CV},
}

@Article{2017arXiv170400028G,
  author        = {{Gulrajani}, Ishaan and {Ahmed}, Faruk and {Arjovsky}, Martin and {Dumoulin}, Vincent and {Courville}, Aaron},
  title         = {{Improved Training of Wasserstein GANs}},
  journal       = {arXiv e-prints},
  year          = {2017},
  pages         = {arXiv:1704.00028},
  month         = mar,
  adsurl        = {https://ui.adsabs.harvard.edu/abs/2017arXiv170400028G},
  archiveprefix = {arXiv},
  doi           = {10.48550/arXiv.1704.00028},
  eid           = {arXiv:1704.00028},
  eprint        = {1704.00028},
  primaryclass  = {cs.LG},
}

@Article{ma_quantum_2025,
  author   = {Ma, QuanGong and Hao, ChaoLong and Si, NianWen and Chen, Geng and Zhang, Jiale and Qu, Dan},
  title    = {Quantum adversarial generation of high-resolution images},
  journal  = {EPJ Quantum Technology},
  year     = {2025},
  volume   = {12},
  number   = {1},
  pages    = {3},
  month    = jan,
  issn     = {2196-0763},
  doi      = {10.1140/epjqt/s40507-024-00304-3},
  url      = {https://doi.org/10.1140/epjqt/s40507-024-00304-3},
}

@ARTICLE{2003quant.ph…1079L,
       author = {{Lavor}, C. and {Manssur}, L.~R.~U. and {Portugal}, R.},
        title = "{Grover's Algorithm: Quantum Database Search}",
      journal = {arXiv e-prints}, 
         year = 2003,
        month = jan, 
          eid = {quant-ph/0301079}, 
        pages = {quant-ph/0301079},
          doi = {10.48550/arXiv.quant-ph/0301079},
archivePrefix = {arXiv}, 
       eprint = {quant-ph/0301079},
 primaryClass = {quant-ph},
       adsurl = {https://ui.adsabs.harvard.edu/abs/2003quant.ph..1079L}
}

@Article{2021arXiv211101118K,
  author        = {{Kang}, Minguk and {Shim}, Woohyeon and {Cho}, Minsu and {Park}, Jaesik},
  title         = {{Rebooting ACGAN: Auxiliary Classifier GANs with Stable Training}},
  journal       = {arXiv e-prints},
  year          = {2021},
  pages         = {arXiv:2111.01118},
  month         = nov,
  adsurl        = {https://ui.adsabs.harvard.edu/abs/2021arXiv211101118K},
  archiveprefix = {arXiv},
  doi           = {10.48550/arXiv.2111.01118},
  eid           = {arXiv:2111.01118},
  eprint        = {2111.01118},
  primaryclass  = {cs.CV},
}

@ARTICLE{9770060,
  author={Yun, Ilwi and Lee, Hyuk-Jae and Rhee, Chae Eun},
  journal={IEEE Transactions on Circuits and Systems for Video Technology}, 
  title={AAGAN: Accuracy-Aware Generative Adversarial Network for Supervised Tasks}, 
  year={2022},
  volume={32},
  number={10},
  pages={6573-6583},
  doi={10.1109/TCSVT.2022.3172998}}

@ARTICLE{2021arXiv211106849J,
       author = {{Jiang}, Liming and {Dai}, Bo and {Wu}, Wayne and {Change Loy}, Chen},
        title = "{Deceive D: Adaptive Pseudo Augmentation for GAN Training with Limited Data}",
      journal = {arXiv e-prints},
         year = 2021,
        month = nov,
          eid = {arXiv:2111.06849},
        pages = {arXiv:2111.06849},
          doi = {10.48550/arXiv.2111.06849},
archivePrefix = {arXiv},
       eprint = {2111.06849},
 primaryClass = {cs.CV},
       adsurl = {https://ui.adsabs.harvard.edu/abs/2021arXiv211106849J}
}

@Article{2025arXiv251207253X,
  author        = {{Xu}, Handing and {Nie}, Zhenguo and {Peng}, Tairan and {Pan}, Huimin and {Liu}, Xin-Jun},
  title         = {{DGGAN: Degradation Guided Generative Adversarial Network for Real-time Endoscopic Video Enhancement}},
  journal       = {arXiv e-prints},
  year          = {2025},
  pages         = {arXiv:2512.07253},
  month         = dec,
  adsurl        = {https://ui.adsabs.harvard.edu/abs/2025arXiv251207253X},
  archiveprefix = {arXiv},
  eid           = {arXiv:2512.07253},
  eprint        = {2512.07253},
  primaryclass  = {cs.CV},
}

@Article{2024arXiv240201791S,
  author        = {{Shu}, Runqiu and {Xu}, Xusheng and {Yung}, Man-Hong and {Cui}, Wei},
  title         = {{Variational Quantum Circuits Enhanced Generative Adversarial Network}},
  journal       = {arXiv e-prints},
  year          = {2024},
  pages         = {arXiv:2402.01791},
  month         = feb,
  adsurl        = {https://ui.adsabs.harvard.edu/abs/2024arXiv240201791S},
  archiveprefix = {arXiv},
  doi           = {10.48550/arXiv.2402.01791},
  eid           = {arXiv:2402.01791},
  eprint        = {2402.01791},
  primaryclass  = {quant-ph},
}

\end{document}